\documentclass[10pt, conference]{IEEEtran}
\makeatletter
\def\ps@headings{%
\def\@oddhead{\mbox{}\scriptsize\rightmark \hfil \thepage}%
\def\@evenhead{\scriptsize\thepage \hfil \leftmark\mbox{}}%
\def\@oddfoot{}%
\def\@evenfoot{}}
\makeatother
\usepackage{array}% http://ctan.org/pkg/array

\usepackage{multirow}
\usepackage{longtable}

\usepackage{indentfirst}

\usepackage{cite}
\ifCLASSINFOpdf
   \usepackage[pdftex]{graphicx}
   \usepackage{epstopdf}
   \usepackage{float}

\else

   \usepackage[dvips]{graphicx}

\fi

\usepackage{bbding}
\usepackage{algorithmic}
\usepackage{algorithm}
\usepackage{amsmath}
\usepackage{amssymb}

\usepackage{array}

\usepackage{bm}

\usepackage{comment}

\usepackage{listings}

\let\labelindent\relax
\usepackage{enumitem}

\usepackage[tight,footnotesize]{subfigure}
\usepackage[bottom=1.03 in,top=0.75in,left=0.635in,right=0.64in]{geometry}

\title{An Accountable Anonymous Data Aggregation Scheme for Internet of Things}

 \author{
 \IEEEauthorblockN{Longfei Wu\textsuperscript{$^{1}$}, Xiaojiang Du\textsuperscript{$^{2}$}, Jie Wu\textsuperscript{$^{2}$}, Jingwei Liu\textsuperscript{$^{3}$}, and Eduard C. Dragut\textsuperscript{$^{2}$}}

 \IEEEauthorblockA{
\textsuperscript{$^{1}$}Department of Mathematics and Computer Science, Fayetteville State University, Fayetteville, NC, USA, 28301\\}

\textsuperscript{$^{2}$}Department of Computer and Information Science, Temple University, Philadelphia, PA, USA, 19122\\

 \textsuperscript{$^{3}$}State Key Lab. of Integrated Services Networks, Xidian University, Shaanxi, Xi’an, China, 710071\\

 \textsuperscript{} Email: \textsuperscript{$^{1}$}lwu@uncfsu.edu,\textsuperscript{$^{2}$}\{dux, jiewu, edragut\}@temple.edu, \textsuperscript{$^{3}$}jwliu@mail.xidian.edu.cn\\
 }

\begin{document}
% make the title area
\maketitle

\begin{abstract}

The Internet of Things (IoT) has become increasingly popular in people's daily lives. The pervasive IoT devices are encouraged to share data with each other in order to better serve the users. However, users are reluctant to share sensitive data due to privacy concerns. In this paper, we study the anonymous data aggregation for the IoT system, in which the IoT company servers, though not fully trustworthy, are used to assist the aggregation. We propose an efficient and accountable aggregation scheme that can preserve the data anonymity. We analyze the communication and computation overheads of the proposed scheme, and evaluate the total execution time and the per-user communication overhead with extensive simulations. The results show that our scheme is more efficient than the previous peer-shuffle protocol, especially for data aggregation from multiple providers.

%%deviation
%%N-2 all collaborative anonymization

%%\boldmath
%
%Despite the increasing capabilities, mobile devices still cannot
%satisfy the computation requirement of many applications.
%Intuitively, this can be solved by outsourcing tasks to external
%resources such as a remote server, cloud, or closely deployed
%cloudlet. However, all of them require extra infrastructures. In
%this paper, we consider a proximate-mobile-device based
%communication system in which all tasks and resources are under the
%control of a central scheduler. We propose a friendship-based task scheduling algorithm to address the contentions when resources are not sufficient. We also present two
%attack models including the denial-of-service (DoS) attack and the
%collusion attack. We evaluate the performance of the proposed
%algorithm along with another contribution-based task scheduling algorithm through extensive experiments.

\end{abstract}
%\vspace{-1 mm}

%\IEEEpeerreviewmaketitle

%%%%%%%%%%%%%%%%%%%%%%%%%%%%%%%%%%%%%%%%%%%%%%%%%%%%%%%%%%%%%%%%%%%%%%%%%%%%%%%%

%\section{WiFi Sharing for Mobile Devices}

%%\section{Note}
%%
%%
%%we consider data same length
%%
%%
%%\textbf{Basic Subsection} lists primitives and the measurement of anonymity. Assume homogeneous \cite{metric1}. \cite{metric2}
%%
%%\textbf{Requirement Subsection} lists the requirements according to the related work's limitation.
%%
%%
%%IoT subgroup does not affect anonymity - users with the same D0j won't be divided
%%
%%
%%how provider server different from 3rd-party server: curious and can identify

%\vspace{-1.5 mm}

\section{Introduction}

Data aggregation is a common task in modern computing systems, such as crowd-sourcing, sensor networks, and cloud computing. However, users are concerned with the privacy of sensitive data, such as medical records, or data that is geo-temporal tagged. 
To motivate the user's active involvement, many anonymous data aggregation schemes have been proposed to preserve data privacy. Threats that may expose the data ownership include the untrusted aggregator, unsecured channels and colluding adversarial participants. The receiver of a packet, like the data collector or an intermediate processor, is capable of tracing back to its immediate sender (e.g., via the source IP address). 
To preserve data anonymity, the aggregation protocol must break the link between a piece of data and its originator (not necessarily the immediate sender).

\begin{figure}[t]
\centering
\includegraphics[width=3.4 in,height=2.1in]{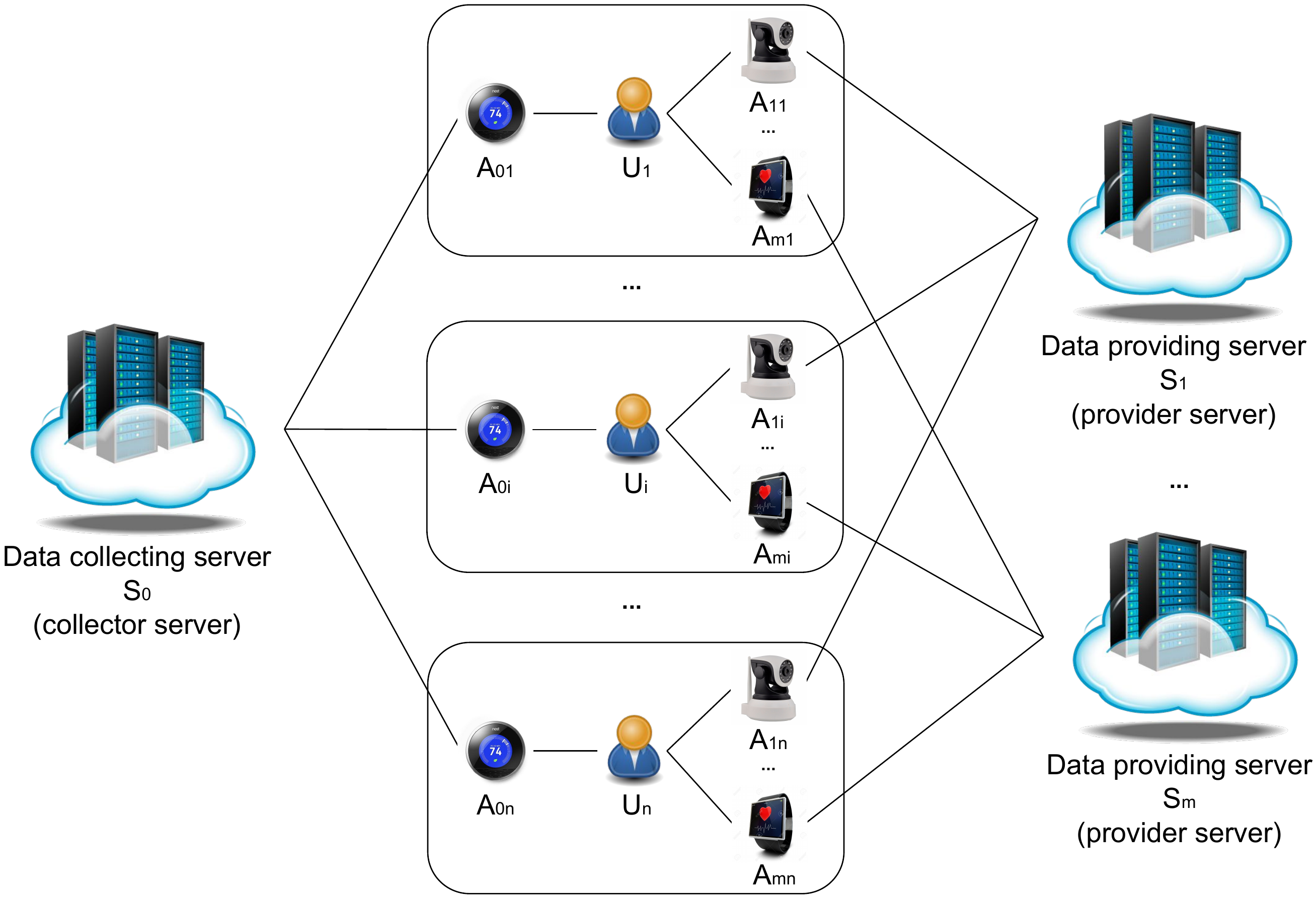}
\vspace{-1mm}
\caption{The System Architecture}
\label{fig:architecture}
\vspace{-4mm}
\end{figure}

Existing anonymous data aggregation mechanisms are designed for the traditional server-client architecture, in which a set of users (clients) submit their data to the collector (server).
However, it is meaningless to discuss data anonymity for the simple server-client data aggregation in IoT ecosystem. In IoT scenario, each IoT company is viewed as the combination of its central server, and a set of client devices deployed with users. 
% As different IoT companies and individual users all sit in different trust domains, the model of security and privacy concerns in the IoT ecosystem are much more complicated than the simple server-client case.
The client devices are the actual generators of data, and provide data readings to their owners.
However, client devices are manufactured by the respective IoT companies, and users do not have the control to prevent the data from being secretly sent to the IoT company server in the background.
% commented in Infocom
%This means that issues like whether the data is to be collected from the client devices and how it is performed are all determined by the IoT companies. 
%There is no way for a user to hide her data from the IoT company if the company wants it, not to mention the fact that most IoT devices, such as medical and monitoring devices, are meant to report the data or be open accessible to the company's server.
%
%
Therefore, instead of studying the anonymity of data aggregation from the IoT devices of an individual company, we focus on data aggregation across multiple IoT companies. This problem can be described using a server-clients-servers framework (shown in Figure~\ref{fig:architecture}), in which multiple client IoT devices are deployed within a home environment. One IoT company (collector server) wants to collect a group of data produced by the client devices of other IoT companies for aggregation studies. For example, a smart home solution company may anonymously collect and analyze the energy consumption and generation data from the residents of a neighborhood, to devise better planning for both energy suppliers and home owners. 
The major difference to the traditional data aggregation is that the collector requires the data from multiple provider servers to be submitted as a group (tuple), while users need to prevent an IoT server (either the collector server or a provider server) from revealing the owner of the data which does not belong to that IoT company. 
%%%%(be it the exact value or whether two users have the equal value).
%
%%%%%%Info Obviously, the existing solutions for single server-clients architecture cannot address the new challenge in IoT system.

In this paper, we study the anonymous data aggregation across multiple providers in the IoT system. The adversaries considered include compromised users, malicious IoT servers, and a global passive eavesdropper. Two passive collusion attacks and 12 active attacks are presented. We design a robust, efficient and accountable aggregation scheme that is resistant to all types of manipulations and collusions by the adversarial entities. We analyze how our scheme can defeat these attacks and provide the accountability, as well as its communication and computation overheads.
The performance is evaluated via simulation tests, and is compared with the Dissent protocol \cite{Dissent}. The results indicate that our scheme is more efficient when collecting data from multiple providers.

%% In this paper, we study the anonymous data aggregation across multiple providers in the IoT system. The adversaries considered include compromised users, malicious IoT servers, and a global passive eavesdropper. Two passive collusion attacks and 12 active attacks are presented. We design a robust, efficient and accountable aggregation scheme that is resistant to all these attacks. We analyze how our scheme can provide the accountability, as well as its communication and computation overheads. The performance is evaluated via simulation tests, and is compared with the Dissent protocol \cite{Dissent}. The results indicate that our scheme is more efficient when collecting data from multiple providers.

%\noindent

Our main contributions are listed as follows:
%\vspace{-0.5 mm}
\begin{itemize}[leftmargin=*]%[nolistsep]
\item To the best of our knowledge, this is the first work that studies the anonymous data aggregation across multiple IoT companies in the server-users-servers architecture of the IoT system, which cannot be properly solved by existing solutions that focus on the traditional server-users structure. 
% To the best of our knowledge, this is the first work that studies the anonymous data aggregation across multiple IoT companies in the server-users-servers architecture.
%
\item Our aggregation scheme does not rely on any trusted 3rd-party server. Instead, it takes advantage of the semi-trustworthy provider servers, considering an IoT company will not leak the data produced by its own client devices to other companies. 
% Though they are not fully trustworthy, our scheme can effectively prevent them from inferring data of other IoT companies or tampering data submission messages.
%
\item We consider a more complex adversary model. Two advanced passive collusion attacks and twelve active attacks specific to our scheme are presented.
\item We analyze the communication and computation overheads, and measure the execution time and the per-user communication overhead with extensive simulations.
%is fully automatic which either need user to make the final decision, or mistakenly accuse benign apps.
%In contrast, precious works either or
\end{itemize}

%\vspace{-1 mm}

%while the users cannot change the rule unless the IoT devices are designed to be programmable. 
% since the IoT client devices (where the data originate) are made by the manufacturer and users do not have full control.

% Although a few IoT devices may give the users the option to only allow access via local network (e.g. WiFi), we follow the broad concept of IoT, which means to let the devices be connected by the Internet so as to integrate of the physical world into computer-based systems. 

% rather than the conventional server-clients data aggregation

%\textbf{commented but maybe useful}

\section{Related Work}

The existing data aggregation schemes can be generally classified into two categories: 
(1) the collector only knows the originator of each encrypted message, but cannot directly decrypt any of the individual messages. Instead, the plaintext is derived through a collaborative computation over a set of ciphertexts, in which the collector cannot figure out the actual contributors of the plaintext data.
(2) the collector receives unencrypted messages that contain the plaintext data, but the immediate sender of each message is not the originator. This happens when the packets have been routed through a set of relays or have been randomly shuffled.

The first class of methodologies include DC-nets \cite{dc-net1988} and $(t; N)$-secret sharing based scheme \cite{SS-2}. 
DC-nets utilize pairwise secrets to conceal the originator of data. 
However, there are two major problems in DC-nets: (i) only one member can embed her data into the ciphertext per slot, while all other members still have to perform the XOR operations (ii) pairwise shared secret keys are required between every two members, and the secrets have to be updated every slot. The generation and distribution of the pairwise secrets can be a heavy burden, especially when there is a large number of members. 
In \cite{SS-2}, one user can divide her secret $s$ into N shares and submitted by N users separately. However, only one user can submit per slot while all others have to generate and share N shares of fake data, which brings large communication overhead.
%%%%%%Info On the collector end, at least $t$ shares are required to reconstruct $s$.

The second class of aggregation schemes consists of Mix-Net \cite{mix1981}, onion routing \cite{OR1998}\cite{Tor}, and peer-shuffle based approach \cite{KDD2006}. 
Mix-Net offers anonymity by relaying encrypted messages through a chain of proxy servers (mixes), which take turns to perform decryption, shuffle, and re-ordering on the messages. 
In onion networks, the encrypted message is transmitted along a path of pre-selected onion routers, each peels away one layer of encryption and uncovers the successor router. 
However, both Mix-Net and early onion routing systems all require reliable and trustful 3rd-party proxies/routers, which may not be available in practice.
The evolved onion routing based systems propose peer-forwarding strategy \cite{Crowds}, in which users themselves serve as the relay for each other. However, the predecessor of the first colluder has a greater chance to be the data originator than other honest users, and this scheme is unable to defend against traffic analysis attacks launched by a global eavesdropper. 
RAC protocol \cite{RAC} solves these issues by enforcing periodical broadcasts for all users, which bring in huge communication overheads.

The peer-shuffle based approach \cite{KDD2006} requires each member to perform shuffle and one layer of decryption on the messages that have been encrypted with all members' public keys. Dissent \cite{Dissent} enhances this protocol to offer accountability against traffic analysis and compromised members.
Then, in order to reduce the heavy computation overheads, many of the above schemes have been modified to be more scalable. Two types of techniques are commonly used. The first type is to gain better scalability while sacrificing anonymity (e.g. reducing anonymity set)\cite{dc-net-3}\cite{k-anonymity}.
The second type utilizes trusted 3rd-party servers to offload the communication and computation overheads \cite{OSDI2012,Verdict,SS-1}. 
Instead, our scheme preserves the maximal anonymity, and does not utilize the assistance of the trustworthy external 3rd-party servers.

A recent work \cite{IoT} studies the anonymous data reporting for participatory sensing in IoT. They adopt the peer-shuffle for slot reservation and pairwise secret XOR for data submission. They still consider data aggregation in the traditional single server-clients structure, which as we have explained, is not a practical issue in IoT system. They do not consider the accountability for attacks and misbehaviors either.

Key management \cite{management1, management2, management3} is essential for security. Several papers \cite{other1,other2,GuanIoTJ,other3,other4,GuanCommMaga,other5} have studied related security issues.

\section{Data Aggregation in IoT}

\subsection{Motivation}

%%%%%%Info In this work, we explore the anonymous data aggregation across multiple IoT companies. A practical application is ``aggregate study''
Aggregate study is a typical application of anonymous data aggregation. In IoT context, one IoT company may need to collect the data of other IoT companies along with its own data, from a certain group of users. By analyzing the patterns and correlations of these data, e.g. using data mining techniques, the collector company can gain better understanding of its data and product. 
In general, an aggregate study is not a strict real-time task, instead is conducted with a relatively high frequency and over a long period of time. This means that the aggregation can bear rather longer elapse caused by computations, but require low communication overhead. 
%, large communication overhead would consume too much bandwidth and may influence regular communications). 
%
Furthermore, since users may not trust either the IoT company servers or a 3rd-party server, the anonymity is achieved via user cooperations only. 
The IoT provider servers may assist the data submission, but only for improving efficiency.
Hence, we adopt a modified version of peer-shuffle technique in our scheme. 
The peer-shuffle can provide anonymity for the messages/data of honest users if no more than $n-2$ out of the $n$ users collude. 
In fact, all sorts of anonymizations based on peer collaboration will fail if $n-1$ users are compromised, so we assume at least two users are honest.
Moreover, considering various adversaries may try to break the data privacy or tamper the data submission, the scheme must be accountable for attacks and misbehaviors.

\subsection{The System Model}

Our system model consists of a set of IoT companies $W = \{W_{0}, W_{1}, ..., W_{m}\}$ and a set of users $U = \{U_{1}, U_{2}, ..., U_{n}\}$. 
An IoT company $W_{i}$ includes a central server $S_{i}$ and a number of affiliated client devices $A_{ij}$ (owned by user $U_{j}, 1\leq j \leq n$). We use $S_{0}$ to denote the data collecting server (collector server), and the rest of the servers $S_{1}, S_{2}, ..., S_{m}$ are the data providing servers (provider servers). All servers maintain connections with the affiliated client devices.
Each user $U_{j}$ is also connected to all IoT client devices in his/her home, $A_{ij} (0\leq i \leq m)$, through a management device (e.g. smartphone). The management device has access to the data generated by IoT devices, and will perform the heavy computations for data anonymization. 
Without loss of generality, we assume that every user owns exactly one client device from each IoT company.
The data reading $D_{ij}$ of device $A_{ij}$ is accessible by both its owner $U_{j}$ and the server $S_{i}$. Additionally, data generated by the same type of client devices (i.e., belonging to the same IoT company) are of the same format and length.
%%%, and maintain stable connection to all devices via a local network (e.g. home WiFi network).
%
%
The collector company $W_{0}$ can aggregate data $D_{ij}$ from multiple provider companies $W_{i} (1\leq i \leq m)$, if approved by the data owners $U_{j}$.
Figure~\ref{fig:architecture} illustrates the many-to-one data aggregation in the server-users-servers architecture. 
%
%
%%%In general, IoT companies can share their users' data if approved by the users. However, they are not permitted to share the data obtained from users behind their back, particularly for data that contains private and sensitive information. In our proposed data sharing scheme, 
%%%%%%Info Users are actively involved not only for granting the data-sharing permission, but more importantly, to jointly perform peer-shuffle so that the collector company $W_{0}$ can learn nothing specific regarding the data from other IoT companies $D_{ij} (1\leq i \leq m)$. 
%
%%%%%%Info Table \ref{table:notations} summarizes the notations used in the system model.

%%%%%%Each IoT company $W_{i} (0\leq i \leq m)$ owns a pair of public/private keys $(KU_{i}^{S}, KR_{i}^{S})$, which are used to encrypt the traffic going in and out of the server. 
%%%%%%%
%%%%%%%
%%%%%%Each user $U_{j} (1\leq j \leq n)$ also possesses a public/private key pair $(KU_{j}^{U}, KR_{j}^{U})$. All public keys are publicly available. 
%%%%%%%
%%%%%%We use $C = E_{KU}[M]$ to denote the encryption of a plaintext $M$ with a public key $KU$, and use $M = D_{KR}[C]$ to denote the decryption of the ciphertext $C$ with the private key $KR$. 

Each IoT company $W_{i} (0\leq i \leq m)$ owns a pair of public/private keys $(KU_{i}^{S}, KR_{i}^{S})$. 
Each user $U_{j} (1\leq j \leq n)$ also possesses a public/private key pair $(KU_{j}^{U}, KR_{j}^{U})$. 
% commented in Infocom
%%All public keys are publicly available. 
%
%We use $C = E_{KU}[M]$ to denote the encryption of a plaintext $M$ with a public key $KU$, and use $M = D_{KR}[C]$ to denote the decryption of the ciphertext $C$ with the private key $KR$. 
%
%
%
%In certain circumstances (e.g. peer-shuffle), we use a different public key encryption algorithm which additionally takes some random bits $R$ as input. 
%
Given a plaintext $M$ and the public key $KU$, the encryption also takes some random bits $R$ as input. 
The produced ciphertext is expressed as $C = {E}_{KU}^{R}[M]$. The plaintext can be recovered with the private key $KR$ ($R$ is not required), $M = {D}_{KR}[C]$.
The advantages of introducing random bits are two-fold: (1) the same piece of plaintext can generate different ciphertexts, so that the equality of data will not be exposed after encryption; (2) the recovery of encrypted data from the decrypted data is disabled.
%
%Various algorithms can achieve this characteristic, including the probabilistic encryption methods like Elgamal \cite{Elgamal} or plaintext expansion/padding. 
Various probabilistic encryption methods can achieve these characteristics, such as the Elgamal encryption algorithm \cite{Elgamal} and the plaintext expansion/padding.
%
%%%%info2  The padding approach is preferred because during the serial encryptions of the peer-shuffle process, plaintext messages are encrypted with all users' public keys and the ciphertext length may increase exponentially to the total number of encryption rounds if using Elgamal encryption, while the padding approach can maintain a ciphertext length of $|M|+O(N)$ which is a linear increase.
%
In peer-shuffle process, the plaintext messages need to be encrypted with all users' public keys (i.e., serial encryptions). 
For the Elgamal encryption algorithm, the ciphertext length may increase exponentially to the total number of encryption rounds $N$.
In contrast, the padding approach can maintain a ciphertext length of $|M|+O(N)$, which is a linear increase.
The plaintext padding can simply extend the plaintext in a predefined format (e.g. $\{R  \enspace ||  \enspace M\}$), or use more advanced and secure schemes like Optimal Asymmetric Encryption Padding (OAEP) \cite{OAEP}. We choose the OAEP for plaintext padding in our scheme.
%
%%%%and the decrypter will just discard the padding random bits $R$ and recover the original plaintext $M$
%
%
%%%\textbf{modulus, http://m.oschina.net/blog/309771}
%%%\textbf{Optimal Asymmetric Encryption Padding (OAEP), %%%https://en.wikipedia.org/wiki/Optimal_asymmetric_encryption_padding, 
%%%https://docs.oracle.com/javase/7/docs/api/javax/crypto/Cipher.html，
%%%http://web.townsendsecurity.com/bid/29195/How-Much-Data-Can-You-Encrypt-with-RSA-Keys$}
%%%cannot remain the same length, change x'(\textbf{after multiple round}), x relation
%
The expression of serial encryptions is abbreviated as
\begin{equation*}
{E}_{KU_{1}:KU_{q}}^{R_{1}:R_{q}}[M] = {E}_{KU_{q}}^{R_{q}} [ {E}_{KU_{q-1}}^{R_{q-1}} [ ... [ {E}_{KU_{1}}^{R_{1}} [M] ]...] ]
\end {equation*}

%each IoT company $W_{i}$ has another pair of public/private keys for probabilistic encryption (e.g. ElGamal encryption) $(PKU_{i}^{S}, PKR_{i}^{S})$, which is used only when it is the data collector. 
%%%%%%Note that $EKU_{i}^{S}$ is actually a key set of three elements, it is also publicly available. The encryption using ElGamal algorithm can be expressed as $\{C_{1}, C_{2}\} = E_{EKU}^{R}[M]$, where $R$ is a random number, $C_{1}$ and $C_{2}$ are the two encrypted messages. The IoT company can compute the plaintext using the private key ($R$ is not required), $M = D_{EKR}[\{C_{1}, C_{2}\}]$.
%The probabilistic encryption algorithm takes the public key $PKU$, a plaintext $M$, and some random bits $R$ as inputs, and produces the ciphertext $C = E_{PKU}^{R}[M]$. The IoT company can compute the plaintext using the private key ($R$ is not required), $M = D_{RKR}[C]$.
%%
%The characteristic of probabilistic encryption algorithm is that the same plaintext can generate different ciphertexts using different random number $R$.
%%
%Similarly, each user $U_{j}$ has an individual public/private key pair for probabilistic encryption $(PKU_{j}^{U}, PKR_{j}^{U})$ as well. The multiple round probabilistic encryption is abbreviated as
%%
%%
%\begin{equation*}
%E_{PKU_{1}:PKU_{q}}^{R_{1}:R_{q}}[M] = E_{PKU_{q}}^{R_{q}} [ E_{PKU_{q-1}}^{R_{q-1}} [ ... [ E_{PKU_{1}}^{R_{1}} [M] ]...] ]
%\end {equation*}

Besides, each IoT company $W_{i} (0\leq i \leq m)$ has another pair of signing/verification keys $(KS_{i}^{S}, KV_{i}^{S})$ to sign a message $M$ as $SIG_{KS_{i}^{S}}[M]$, and verify a signature $sig$ as $VF_{KV_{i}^{S}}[sig]$. 
Similarly, each user $U_{j} (1\leq j \leq n)$ also has the key pair $(KS_{j}^{U}, KV_{j}^{U})$ for signature generation and verification.
We define $\{M\}SIG_{KS}$ as the concatenation (``$||$'') of message $M$ and its signature (signed by key $KS$)
\begin{equation*}
\{M\}SIG_{KS} = M \thickspace || \thickspace SIG_{KS}[M]
\end {equation*}
The hash value of a message $M$ is denoted as $Hash[M]$.
The permutation function $p(z)$ randomly shuffles a group of objects, and assigns a new position for the $z$th object.
The pseudorandom function $PRF(L, seed)$ is used to generate the most significant $L$ bits from $seed$. The random numbers used in encryptions can be generated using this function.

\subsection{Referenced Data Aggregation}

To perform an aggregate study, the collector server needs to gather a set of data from the users of interest. For user $U_{j}$, the data to be submitted includes the readings of all provider companies' devices (provider data). We define this set of data as the provider data set $P_{j} = \{D_{ij}\}, 1\leq i \leq m$. 
The provider data set $P_{j}$ and the reading of the collector company's device (collector data) compose a data tuple $T_{j}$, $T_{j} = \{D_{0j}, P_{j}$\}. 
%
%%%%%%Info Data tuple is the basic unit for aggregate study. 
%%%%%%Info The collector can investigate the correlation between the collector data $D_{0j}$ and the provider data set $P_{j}$ ($1\leq j \leq n$) through a number of data tuples over time.
%
The traditional data aggregation is one-dimensional, hence users only submit their independent pieces of data. However, the data aggregation in the IoT is two-dimensional. Each user $U_{j}$ needs to submit the provider data set $P_{j}$, which is referenced by the collector data $D_{0j}$. 
Since the collector server knows $D_{0j}$ for each user, it can find the owner of data tuple $T_{j}$ if the $D_{0j}$ is unique among all users.
Hence, it should avoid collecting from a user with unique $D_{0j}$, for example, by splitting the data range into small segments and rounding $D_{0j}$ values up/down to the nearest segment. 
%
%%%%%%Info Intuitively, we need a uniqueness checking to guarantee that no data tuple will be exposed. 
%
% commented for Infocom
%%%%%%Info In general, the probability of uniqueness arises when the aggregation set is small. In this case, the collector can split the data range into segments of equal size, and round the $D_{0j}$ values up/down to the nearest segment. 
%
%%%%%%Info Intuitively, we need a uniqueness checking in case the collector intentionally requests for data tuple with unique $D_{0j}$.
%%%%%%Info In fact, the collector should take the responsibility to check the uniqueness before sending out aggregation requests.
%
%
%%% For simplicity, we assume there is no unique $D_{0j}$.
%%%%%%Info But still, users have to perform uniqueness checking to confirm, in case the collector intentionally makes a certain $D_{0j}$ value unique among its requests.
%
%%%Besides, it is useless to anonymize data tuples among all users - the collector can narrow the range of the originator by comparing the $D_{0j}$.
%%%In fact, the aggregation of referenced data intrinsically divide the users into different anonymity sets.
%%%%
%%%Our scheme only anonymizes data tuples with the same $D_{0j}$ value (within each anonymity set).

Another characteristic in IoT data aggregation is the provider servers. Although they are not fully reliable, our scheme takes advantage of these provider servers to delegate the submission of their own data.
%%%%%%Info  while isolating them from the data of other providers. 
%%%%%%Info Specifically, they are not supposed to participate any computation or checking procedure that involves other companies' data, but it is safe to let each provider server $S_{i} (1\leq i \leq m)$ submit its own data $D_{ij}$ for their users $U_{j} (1\leq j \leq n)$. 
If users submit the provider data set by themselves, they need to perform peer-shuffle on the data in order to prevent being traced  by the collector. All such efforts can be saved if each corresponding provider server submits the provider data for its users. 

Besides, considering that the receiver may get an updated data reading by the time the request arrives (which is not the data intended by the requester), we need to solve the inconsistency caused by the elapse of message transmissions. Specifically, the data $D_{0j}$ is embedded in the aggregation request sent to users, and the data $D_{ij}$ is embedded in the submission request sent to provider servers.

%%%%%%Info Besides, considering the whole IoT system may not be well-synchronized and the elapse of message transmissions, the inconsistency between the provider data intended by the collector and the data actually aggregated from the providers must be avoided. For example, the user or provider server may have obtained updated data reading by the time the request arrives, and as a result, submit the new reading instead.
%%%%%%Info For instance, the collector aims to diagnose an abnormal data $D_{0j}$ with the corresponding provider data set $P_{j}$ (measured at the same time with $D_{0j}$), while the IoT devices may have obtained updated data readings  by the time the requests arrive and submit the new readings instead.
%
%%%%%%Info To satisfy the collector's rigid timing requirement for the data to be aggregated, the data $D_{0j}$ is embedded in the aggregation request sent to users, and the data $D_{ij}$ is embedded in the submission request sent to provider servers.

%\vspace{-1mm}

\subsection{The Adversary Model}
\label{adversary}

%%%The users are often unwilling to provide sensitive data such as medical records and accurate location. They are willing to share these information only if they are assured that the data will be used solely for aggregate study purpose and that the data cannot be linked back to them. Hence, the submission of data to the collector company has to be anonymous. 
%
%\vspace{-0.5mm}
\subsubsection{IoT Companies and Users}
All IoT companies and users are dedicated to preserving the privacy of their own data, meanwhile they are curious about other's data. 
We assume that all IoT companies will obey the commitment not to disclose their user's data without permission, so the IoT company that deliberately reveal or trade data to other IoT companies is not considered within the scope of this paper. %%%%%%Info This means that they will not share data with each other, or expose data to unauthorized users.
Additionally, the client devices are the source of data, and they will always provide the true data to their owners and the corresponding company servers. 
%
%
%%%%%%Info The curiosity inclination of IoT servers and users drives them to illegally gain data that they are not supposed to have access. Specifically, the IoT servers are interested in inferring user's data provided by other companies, while users attempt to infer the data of other users.
%
However, malicious users and provider servers may deviate from the protocol.
%%%%%%Info2  interrupt the data aggregation.
% collector server will follow does not interrupt the data aggregation initiated by itself, 
%%%%%%Info2 The malicious provider servers may tamper with the data during submission.
%
%
We consider two basic adversarial actions: \emph{collusion} and \emph{manipulation}. 
Collusion means that some compromised users may cooperate with each other or with the collector/provider server to de-anonymize the data of honest users. Manipulation allows servers and users to insert, delete or tamper the messages handled by them. 
%

%%%%%%Info Furthermore, we classify malicious attacks into three categories based on their purposes. The first category is the disruption attacks, which target at disrupting the aggregation protocol. 
%
%%%%%%Info The second category is the data inference attacks, in which the adversaries either work individually or collude to infer the data of honest users.
%
%%%%%%Info The third category is the data falsification attacks aiming to submit falsified data to spoof the collector.
%
%%%%%%Info The disruption of the protocol and the falsification of data must be made by active attacks, while the data inference attacks can be further divided into passive inference attacks (no manipulation) and active inference attacks.

%\vspace{-0.5mm}
\subsubsection{Eavesdropper}
A global and passive eavesdropper may exist, which can monitor all traffics in the network. It may collude with an IoT server and malicious users. 

\section{The Proposed Scheme}

In this section, we present the proposed accountable anonymous data aggregation scheme. The scheme works in 6 phases, $\sigma$ is the phase ID.
%$(1\leq \sigma \leq 6)$. 
%%%% There is one round of communication in each phase between users and servers, the message transmitted in phase $\sigma$ is denoted by $\Omega_{\sigma}$.
The message transmitted in phase $\sigma$ is denoted by $\Omega_{\sigma}$.
%the message transmitted in phase $\sigma$ is denoted by $\Omega_{uv}^{xy}(\sigma)$, where $x$ and $y$ are the entity type ($S$ for server or $U$ for user) of the sender/receiver, $u$ and $v$ are the entity ID of the sender/receiver.
%
Each run of data aggregation is uniquely identified by the session ID $sid$.
To hold the server and user behaviors accountable, the messages they transmit in each phase are associated with the session ID $sid$ and phase ID $\sigma$, and are signed with the sender's signing key. The receivers will first verify the signature of the received messages before further processing.
All transmitted and received messages as well as the random numbers used for encryptions
%%%%%and the permutations performed in peer-shuffle 
are recorded until the next protocol run, to serve as the evidence in case an investigation of misbehaviors is conducted.

\subsection{The Data Aggregation Protocol}

\noindent\textbf{Phase 1: Aggregation requests}

The aggregation begins at the collector server $S_{0}$, who sends the aggregation requests for data $D_{ij} (1\leq i \leq m)$ to each user $U_{j}$.
%
%
%
%
%%%% All group performs the anonymization and checkings autonomously. 
% for the collector server sending its data aggregation requests to users, and the  for the users sending their submmission requests to the provider servers, the schemes requires the data to be embedded in the request messages. This can avoid the inconsistency if the receiver has obtained newer data by the time the request arrives.
%
%
%
%%%%The collector server $S_{0}$ first divides users into different groups (anonymity sets), based on the $D_{0j} (1\leq j \leq n)$ value. Then, it sends the aggregation requests of data $D_{ij} (1\leq i \leq m)$ to all users.
%%%%% via client device $A_{0j}$. 
%
%%%The request message sent to user $U_{j}$ contains the data $D_{0j}$ and the list of members $l_{j}$ in $j$'s group (both are encrytped using user $j$'s public key $KU_{j}^{U}$).
The request message contains the collector data $D_{0j}$, and it is encrypted with $U_{j}$'s public key $KU_{j}^{U}$ and the random number $R_{0j}$ selected by the collector server.
%
%%%%%%%%%%%%%%%%%Without losing generality, we put $D_{0j}$ into in phase 1 messages as below, but it can be empty when the request is not specified for the data tuple with a particular $D_{0j}$.
%
%
\begin{equation*} 
% \Omega_{1} = \{ E_{KU_{j}^{U}}[D_{0j}, l_{j}], sid, 1 \} SIG_{KS_{0}^{S}}
\Omega_{1} = \{ {E}^{R_{0j}}_{KU_{j}^{U}}[D_{0j}], sid, \sigma_{1} \} SIG_{KS_{0}^{S}}
\end {equation*}
%
%\vspace{-1mm}
%%%notifies the affiliated client devices $A_{0j}$ about its intent to aggregate data $D_{0j} (1\leq i \leq m)$ from each corresponding user $j$. Then, each client device $A_{0j}$ forwards the aggregation request to its owner $j$, along with its data reading $D_{0j}$.
%%%%where  is the random number selected by the collector server $S_{0}$ for $U_{j}$. 

\noindent\textbf{Phase 2: Submission of index messages}

%%%%info2  User approval and 

%%%% Upon receiving the request, users first run the uniqueness checking (Section~\ref{subsec:ASC}). Then, each user $j$ can decide whether to approve or not. If not, user $j$ needs to inform the collector as well as other members in $l_{j}$, who will update the group member list respectively (remove user $j$ from the list). 

%%%%%%%%%%%%%%Upon receiving the aggregation request, users first run the request verification checking (Section~\ref{subsec:RVC}). Then, each user $U_{j}$ can decide whether to approve or not. If not, $U_{j}$ needs to inform all other users as well as the collector server before quit. 

Upon receiving the aggregation request, each user $U_{j}$ prepares a pair of information: the collector data $D_{0j}$ and a pseudonym number $PN_{j}$ (as the index number). $PN_{j}$ is an L-bit random number generated using $PRF(L, seed_{j})$, where $seed_{j}$ is randomly selected by $U_{j}$. The probability that no collision occurs among the selected index numbers is
%%%%%%Info Upon receiving the aggregation request, each user $U_{j}$ can decide whether to approve or not. If not, $U_{j}$ needs to inform all other users as well as the collector server before quit. 
%%%%%%Info 
%%%%%%Info Otherwise, $U_{j}$ prepares a pair of information: the collector data $D_{0j}$ and a pseudonym number $PN_{j}$ (as the index number). $PN_{j}$ is an L-bit random number generated using $PRF(L, seed_{j})$, where $seed_{j}$ is independently and randomly selected by $U_{j}$. The probability that no collision occurs on $PN_{j}$ is
%
\begin{equation*} 
 \frac{\prod_{0\leq a \leq n-1} (2^{L}-a)}{2^{Ln}}
\vspace{-1mm}
\end {equation*}
%
%%%Given a group size $|l_{j}|$, 
$L$ is selected to limit the collision probability under a sufficient small threshold (e.g., $10^{-3}$). 
%
%
%%%%%Next, users encrypt their index numbers with the collector's public key $KU_{0}^{S}$. 
%
%%%\begin{equation}
%%%EPN_{j} = E_{KU_{0}^{S}}[PN_{j}]
%%%\end {equation}
%
The data $D_{0j}$ and index number $PN_{j}$ pair 
%(both encrypted with the collector server's public key $KU_{0}^{S}$) 
is defined as the index message (IM) of $U_{j}$,
\begin{equation*}
IM_{j} =  \thickspace <  \thinspace D_{0j}, PN_{j}  \thinspace >
\vspace{-1.5mm}
\end {equation*}
%
%
%
%
%%% encrypted IM old
%%%Then each user $j$ encrypts her index message $IM_{j}$ with the public keys of all group members in $I_{j}^{G}$, following the order they appear in $l_{j}$ (e.g. numerical order).
%%%Here we look into one group for illustration. We assume this group is composed of user $U_{1},...,U_{a}$. 
%%%%
%%%The encrypted index message is
%%%\begin{equation*}
%%%EIM_{j}(K_{a}) = \{ IM_{j} \}_{KU_{1}^{U}:KU_{a}^{U}}
%%%\end {equation*}
%%%%
%%%where $K_{a}$ is the ordered set of $a$ keys that are sequentially used for encryption. In this case, $K_{a}=\{KU_{1}^{U},KU_{2}^{U},...,KU_{a}^{U}\}$.
%%%%
%%%Note that when all the $a$ folds of encryptions are decrypted, $EIM_{j}(k_{0})=IM_{j}$. 
%%%%
%%%Next, all encrypted index messages are sent to the last group member for individual anonymization processing (IAP).
%%%%
%%%\begin{equation*} 
%%%\Omega_{0j}^{SU}(1) \thickspace = \thickspace \{ D_{0j}, l_{j}, sid \} SIG_{KS_{0}^{S}}
%%%\end {equation*}
%%%
%%% next is new:
%%%%%Then, each user j encrypts her index message IMj with the public keys of all
%
Then, the index message $IM_{j}$ is encrypted using the public keys of all users, following a given order (e.g. sequential order).
%
%%%%%From now on, every group works autonomously, including the index message encryption, anonymization and verification.
For illustration, here we conduct the serial encryptions in the order $U_{1},...,U_{n}$
%%%%%%Info Here we look into an example for illustration, in which all users have approved the aggregation and the serial encryptions are made in the order $U_{1},...,U_{n}$. %%%%We assume this group is composed of user $1$ to $a$, in the order $U_{1},...,U_{a}$. 
%
The encrypted index message (EIM) is 
\begin{equation*}
EIM_{j} = \thickspace  {E}_{KU_{1}^{U}:KU_{n}^{U}}^{R^{\prime}_{j1}:R^{\prime}_{jn}}[IM_{j}] \thinspace
\vspace{-1.5mm}
\end {equation*}
%%%\begin{equation*}
%%%EIM_{j}(K_{a}) = \{ IM_{j} \}_{KU_{1}^{U}:KU_{a}^{U}}
%%%\end {equation*}
%%%%
%%%where $K_{a}$ is the ordered set of $a$ keys that are sequentially used for encryption. In this case, $K_{a}=\{KU_{1}^{U},KU_{2}^{U},...,KU_{a}^{U}\}$.
%%%%
%%%Note that when all the $a$ folds of encryptions are decrypted, $EIM_{j}(k_{0})=IM_{j}$. 
%%%%
where $R^{\prime}_{j1},...,R^{\prime}_{jn}$ are the random numbers selected by $U_{j}$ for the encryptions with public keys $KU_{1}^{U},...,KU_{n}^{U}$, respectively. All encrypted index messages are sent to the first processor for individual anonymization processing (IAP). 
%
%%%%In other words, the first IAP processor is the user who has conducted the last round of encryption.
%
\begin{equation*} 
%%%\Omega_{ja}^{UU}(2) \thickspace = \thickspace \{ EIM_{j}, sid \} SIG_{KS_{j}^{U}}
\Omega_{2} = \{ EIM_{j}, sid, \sigma_{2} \} SIG_{KS_{j}^{U}}
\vspace{-1.5mm}
\end {equation*}
%%%%where $U_{a}$ is the last group member in $l_{j}$.

%\vspace{-2.5mm}

\noindent\textbf{Phase 3: Anonymization of index messages}

The data anonymization is achieved via the IAPs by each user (processor), performed in the reverse order of the serial encryptions.
Before the processors begin the IAP process, they need to make sure that all EIMs are valid: their received EIMs have been properly handled without being ``marked".
For these purposes, each processor carries out the replication attack checking (Section \ref{subsec:replicate}). 
In our example, the first processor is the last encryptor $U_{n}$. The received EIMs are concatenated as an anonymization bundle message (ABM),
\begin{equation*}
ABM_{n} = EIM_{1} \thickspace || \thickspace ... \thickspace || \thickspace EIM_{n}
\vspace{-.5mm}
\end {equation*}
The EIMs can be in any order (e.g., the order they arrive at the processor). For simplicity, here we list them sequentially.
The IAP process contains two steps: shuffle and decryption. User $U_{n}$ first shuffles the EIMs with its random permutation function $p_{n}$,
\begin{equation*}
ABM_{n}  \xrightarrow[p_{n}]{shuffle}  EIM_{p_{n}(1)} \thinspace || \thinspace ... \thinspace || \thinspace EIM_{p_{n}(n)}
\vspace{-.5mm}
\end {equation*}
Then $U_{n}$ decrypts all pieces of EIMs with her private key $KR_{n}^{U}$.
When the IAP finishes, the ABM becomes
%%%the $ABM_{n}$ becomes ($U_{a}$'s piece of EIM is highlighted)
%
\begin{equation*}
ABM_{n}^{\prime} = {D}_{KR_{n}^{U}}[EIM_{p_{n}(1)}] \thinspace || \thinspace ... \thinspace || \thinspace {D}_{KR_{n}^{U}}[EIM_{p_{n}(n)}]
%%%%ABM_{n}^{\prime} =  \thinspace || \thinspace [  \thinspace \{ D_{0a} \}_{EKU_{1}^{U}:EKU_{n-1}^{U}}^{R_{1}:R_{n-1}}, \{ PN_{a} \}_{KU_{1}^{U}:KU_{n-1}^{U}} \thinspace ]_{p_{n}(a)} \thinspace || \thinspace
%%%EIM_{j}^{\prime} = \thinspace < \{ D_{0j} \}_{EKU_{1}^{U}:EKU_{n-1}^{U}}^{R_{1}:R_{n-1}}, \{ PN_{j} \}_{KU_{1}^{U}:KU_{n-1}^{U}} \thinspace >
\vspace{-.5mm}
\end {equation*}
As all other users repeat the IAP (i.e., each shuffles the EIMs and strips off one layer of encryption), the last processor (i.e., $U_{1}$) will be able to recover the original index messages in a new random order. 
%
%
%\vspace{-1mm}
\begin{equation*}
%%%\begin{split}
%%%EIM(k_{0})
%%%& = EIM_{{p}(1)}(k_{0}) || ... || EIM_{{p}(n)}(k_{0}) \\
%%%& = IM_{{p}(1)} || ... || IM_{{p}(n)}
%%%\end{split}
ABM_{1}^{\prime} = IM_{{p}(1)} \thickspace || \thickspace ... \thickspace || \thickspace IM_{{p}(n)}
\vspace{-1.5mm}
\end{equation*}
where ${p}$ stands for a series of permutations $p_{1},..., p_{n}$ made by each processor.  
The recovered index messages are broadcasted to all users as well as the collector server $S_{0}$.
In phase 3, the transmitted messages are the resulted $ABM^{\prime}$ after each IAP, 
\begin{equation*} 
%%%\Omega_{ja}^{UU}(2) \thickspace = \thickspace \{ EIM_{j}, sid \} SIG_{KS_{j}^{U}}
\Omega_{3} = \{ ABM^{\prime}, sid, \sigma_{3} \} SIG_{KS}
\vspace{-1.5mm}
\end {equation*}
where $KS$ is the signing key of the corresponding processor.
%%%where $ABM^{\prime}$ is the final $ABM$ (permuted index messages) after the completion of anonymization.

\noindent\textbf{Phase 4: Peer-shuffle verification and submission requests}

%At the start of phase 4, all users once again execute three detections (group consistence checking, replication attack checking, and replacement attack checking), considering the case that the last several processors are malicious and they may launch those attacks just to 
%
%Each user $U_{j}$ has to execute three detections: (1) index message checking (Section~\ref{subsec:ASC})

At the start of phase 4, all users first run the replacement attack checking (Section \ref{subsec:replace}) and the broadcast consistence checking (Section \ref{subsec:BCC}), to verify the consistency of the broadcasted final ABM and confirm the existence of their own index messages. 
%
%%%%If passed, it is guaranteed that the IMs of honest users have not been manipulated. 
%
Then, the uniqueness checking (Section \ref{subsec:UC}) is performed to check the existence of unique $D_{0j}$ or identical index numbers among IMs.
After these checkings are passed, each user $U_{j}$ informs the provider servers $S_{i} (1\leq i \leq m)$ to submit the data $D_{ij}$ to the collector for them.
%
% collector's information $Info(S_{0})$ (e.g. server IP address and port number)
Specifically, $U_{j}$ prepares a data submission message $DSM_{ij}$ which contains the data $D_{ij}$ and the index number $PN_{j}$. 
%%%%%%%%%%%%Similar to the aggregation requests, $D_{ij}$ can be empty if the collector just wants to gather the most recent reading (instead of the specific $D_{ij}$ corresponding to a particular $D_{0j}$).
%
The index number is encrypted using the collector's public key $KU_{0}^{S}$ and a random number $R_{j}$. The submission message $DSM_{ij}$ is expressed as
%%%\begin{equation*}
%%%\{C^{1}_{j}, C^{2}_{j}\} = E_{EKU_{0}^{S}}^{R}[ PN_{j} ]
%%%\end {equation*}
%
\begin{equation*} 
DSM_{ij} = < D_{ij}, {E}_{KU_{0}^{S}}^{R_{j}}[ PN_{j} ] >
\vspace{-1.5mm}
\end {equation*}
Finally, the data submission message is encrypted using the provider server's public key $KU_{i}^{S}$ and another random number $R_{j}^{\prime}$, and sent to the provider server $S_{i}$,
\begin{equation*} 
\Omega_{4.2} = \{ {E}_{KU_{i}^{S}}^{R_{j}^{\prime}}[DSM_{ij}], sid, \sigma_{4.2} \} SIG_{KS_{j}^{U}}
\vspace{-1.5mm}
\end {equation*}
\noindent\textbf{Phase 5: Data submission}

%%%%%After decrypting the data submission messages, the provider servers will submit the data and encrypted index numbers to the collector server. 
After receiving the data submission messages from users, provider servers will submit them to the collector altogether.
For a given provider server $S_{i} (1\leq i \leq m)$, the data of all its users are submitted as  
%%%%The data $D_{ij}$ is encrypted with the collector server's public key $KU_{0}^{S}$
\begin{equation*} 
\Omega_{5} = \{ DSM_{i1} || ... || DSM_{in}, sid, \sigma_{5} \} SIG_{KS_{i}^{S}}
\vspace{-1.5mm}
\end {equation*}

%After the provider servers $S_{i} (1\leq i \leq m)$ gather all the requests from the local devices, they extract the $Info(S_{0})$ segment and group the requests according to their collector companies. For each specific collector $S_{0}$, provider servers reply with all the data packets (i.e. $<PN_{j}, D_{ij}>$ pairs) combined together with no specific order. 

%\vspace{-1mm}
\noindent\textbf{Phase 6: Data submission verification}

%
%%%%%The collector server receives each aggregated data in two parts: one part is the $<E_{KU_{0}^{S}}[D_{0j}], E_{KU_{0}^{S}}[PN_{j}]>$ pair (received by the end of phase 3) and the other part is the $<E_{KU_{0}^{S}}[D_{ij}], E_{PKU_{0}^{S}}^{R_{j}}[ PN_{j} ]> (1\leq i \leq m)$ pairs. 
The collector server receives the aggregated data tuples in two separate parts: the first part is the $<D_{0j}, PN_{j}>$ pair (received by the end of phase 3) and the second part is the $<D_{ij}, {E}_{KU_{0}^{S}}^{R_{j}}[ PN_{j} ]>$ pairs. 
%
%%%It first uses the private key $KR_{0}^{S}$ and the probabilistic decryption private key $PKR_{0}^{S}$ to decrypt these pairs, and recover the plaintext data $D_{0j},...,D_{mj}$ and index number $PN_{j}$ for each $U_{j}$.
It first decrypts each ${E}_{KU_{0}^{S}}^{R_{j}}[ PN_{j} ]$ with its private key $KR_{0}^{S}$.
Then, the original data tuple $T_{j}$ can be reconstructed by linking the $<D_{0j}, PN_{j}>$ and $<PN_{j}, D_{ij}>$ pairs via the unique index number $PN_{j}$. 
%
%It can obtain the data pairs $<D_{0j}, D_{ij}>$ from all the users of interests. Since there are at least two users with the same $D_{0j}$ value, the collecctor is unaware of which user a specific pair of data belongs to.  
%
The data submission checking (Section \ref{subsec:DSC}) must be performed by each user, in case the provider servers may have manipulated the submitted data or the index number.
%%However, the authenticity of the provider data set cannot be confirmed since the provider servers may have manipulated the submitted data or the index number. The data submission checking (\ref{subsec:DSC}) is performed to verify that the data tuple has been correctly aggregated by the collector server.
The collector server will finally accept the provider data if the checking is passed.

\noindent\textbf{List of Checkings:}

\subsubsection{Replication Attack Checking}
\label{subsec:replicate}

each processor scans all EIMs, and checks if any two are duplicated or have the same encrypted index number. With extremely low probability, the duplication can be caused by the collision during the index number generation when their owners select the same random numbers for serial encryptions. This can be confirmed by replaying the IAP processes backwards (i.e., re-encrypting the index messages layer-by-layer with corresponding public keys and logged random numbers), and then comparing with the logged original EIMs. We only focus on whether the reconstructed EIMs exist in the logged messages, while the permutations are not replayed. If the duplication is indeed due to the collision, the aggregation procedure will be restarted. Otherwise, it must be a replication attack (Attack 7) launched by compromised processor(s):
the aggregation procedure is aborted, and an investigation is launched to replay the executed IAP processes.
%%%%and find the attacker(s) who have copied the EIM of honest users.
%
%
The detection for a replication attack has to be conducted at each processor, otherwise the malicious processors may bypass the checking by changing the replicated EIMs back to the original ones which are owned by themselves.
%%%%info  during the last IAP (assuming the last processor is also compromised).
%

%%\subsubsection{Group Consistence Checking}
%%\label{subsec:GCC}
%%% attack
%%validate if all received (encrypted) index messages are consistent to be in one group. This checking includes
%%(1) if the number of messages equals $|l_{j}|$; (2) if the (encrypted) $D_{0j}$ values in all messages are the same.
%%%
%%The check (1) detects if any user has joined the wrong group, either misled by the collector (Attack 1), deliberately by herself (Attack 3) or malipulated by a precious IAP processor (Attack 5). 
%%%
%%%%%All previous IAPs are examined to see if a malicious processor has inserted or removed any index messages. If all previous processors are clean, the group member who is missing or the external user who joined the group while is not present in $l_{j}$ has to prove that she is misled by the collector, using the aggregation request she received along with her private key. Otherwise, she is accountable for this.
%%The check (2) detects if there is any distinction of the (encrypted) $D_{0j}$ values within the EIMs. Similarly, it may be caused by the collector (Attack 2), the user (Attack 4) or a precious processor (Attack 6). 
%%%
%%If this checking fails, the aggregation procedure is aborted.
%%The attacker can be located by examining the communication logs between the collector and users, as well as the communication logs between users and processors.

\subsubsection{Replacement Attack Checking}
\label{subsec:replace}

each user scans all recovered index messages in the (broadcasted) final ABM, and confirms the existence of its own index message. If it is missing, there must be malicious processor(s) who have replaced its EIM (Attack 8). 
The aggregation procedure is aborted, and the investigation to find the manipulator is launched by replaying the IAP processes.
%
%%%%Two options are available for the detection of replacement attacks. In the first option, it is performed by each IAP processor in phase 3, while the other option only requires it to be performed once at the beginning of phase 4. The first option can detect replacement attack sooner at the cost of $n$ times of checking complexity. Since our scheme is able to detect the replacement attacks, which will defer the adversaries from launching such attacks (or very rarely), we choose the second option to make the scheme more efficient. 
%
Note that malicious processors cannot recover the replaced legitimate EIMs of honest users once they have been processed by an honest processor, as they know neither the honest processor's private key nor the random numbers used by the honest users.
%commented in Infocom
%%%%info  , as they don't have the private keys of honest processors.

\subsubsection{Broadcast Consistence Checking}
\label{subsec:BCC} all users as well as the collector server must verify that the broadcasted final ABMs $ABM_{f}$ they received are consistent.
Since the broadcast is not in a wireless channel, all receivers have to explicitly exchange and compare the received $ABM_{f}$. To make it more efficient, they may only exchange the hash value $Hash[ABM_{f}]$,
\begin{equation*} 
\Omega_{4.1} = \{ Hash[ABM_{f}], sid, \sigma_{4.1} \} SIG_{KS}
%\vspace{-.5mm}
\end {equation*}
where $KS$ is the signing key of the broadcast receiver.
The hash values are compared by each receiver.
If any of them is different from others, the last processor must have launched the broadcast attack (Attack 9).
The aggregation is aborted.

%%%%\subsubsection{Uniqueness Checking}
%%%%\label{subsec:ASC}
%%%%% attack
%%%%%%%The collector may create a group with single user $j$, so that the $D_{0j}$ is unique and the data tuple $T_{j}$ will be disclosed.
%%%%%
%%%%check if the group size $|l_{j}|$ equals 1 (Attack 8). If so, the user $j$ must quit the aggregation.

\subsubsection{Uniqueness Checking}
\label{subsec:UC}
% attack
%%%The collector may create a group with single user $j$, so that the $D_{0j}$ is unique and the data tuple $T_{j}$ will be disclosed.
%
The uniqueness checking on shuffled index messages consists of two parts. 
In the first part, each user checks if its $D_{0j}$ is unique among all IMs. If so, the collector has launched a unique collector data attack (Attack 10).
%%%%info2  , and $U_{j}$ must quit the aggregation to prevent its data tuple $T_{j}$ being identified by the collector server.
In the second part, each user checks if its index number $PN_{j}$ is not unique among all IMs. If so, it can be due to either the coincidence of index number selection, or the manipulation made by the previous IAP processor who has intentionally replicated an honest user's encrypted index number (not necessarily the data part).
Since this is similar to the replication attack checking, it can be handled in the same way and we do not categorize it as an independent attack. 
%%%%info  Whether it is an attack launched by the previous processor or the last several colluding processors can be decided by replaying the phase 3.

%
\subsubsection{Data Submission Checking}
\label{subsec:DSC}

each user needs to ensure that their DSMs have been correctly submitted by the provider servers, who may have tampered with the data $D_{ij}$ and/or the encrypted index number ${E}_{KU_{0}^{S}}^{R_{j}}[ PN_{j} ]$ (Attack 11). This requires the cooperation of the collector server, who will acknowledge each submission by replying with the signed DSM to the respective provider server $S_{i}$,
\begin{equation*} 
\Omega_{6.1} = \medspace \{ DSM_{ij}, sid, \sigma_{6.1} \} SIG_{KS_{0}^{S}} \medspace 
\vspace{-1mm}
\end {equation*}
After all the acknowledgement messages have arrived at the provider servers, each of them sends the signatures of the acknowledgement messages, to the respective user who owns the unique $DSM_{ij}$ (i.e., the unique ${E}_{KU_{0}^{S}}^{R_{j}}[ PN_{j} ]$),
\begin{equation*} 
\Omega_{6.2} =  \medspace \{ {E}_{KU_{j}^{U}}^{R_{ij}}[ SIG_{KS_{0}^{S}}[DSM_{ij} ]], sid, \sigma_{6.2} \} SIG_{KS_{i}^{S}} \medspace 
\vspace{-1mm}
\end {equation*}
Where $R_{ij}$ is the random number selected by the provider server $S_{i}  (0\leq i \leq m)$ for its user $U_{j}$.
Finally, each user will verify the signature of the acknowledgement message and the contained $DSM$. If failed, the user can directly warn the collector that her data submission has been manipulated by provider $S_{i}$, so that the collector can discard the falsified data.
%
%%%%%%%Note that no matter it is the data $D_{ij}$ (true data not submitted) or the encrypted index number ${E}_{KU_{0}^{S}}^{R_{j}}[ PN_{j} ]$ (the collector cannot link this index number to the correct $<D_{0j}, PN_{j}>$ pair) that is tampered, the collector server cannot know $U_{j}$'s real data tuple $T{j}$.  

\subsection{Scheme-specific Active Attacks}
\label{subsec:attack}

We present various active attacks specific to our scheme, including the disruption attacks, data inference attacks, and data falsification attacks.

\subsubsection{Disruption Attacks} The disruption attacks abort the protocol run by creating an abnormality that it cannot handle. An investigation is conducted immediately when such abnormality happens.
Next, we list the possible disruption attacks launched by the processors and provider servers, separately.
%%%%be launched by all entities, we list them separately based on the attacker identity. 

%%%%info2  \begin{itemize}[leftmargin=*]%[nolistsep]
%%%%%\vspace{-1mm}
%%%%%
%%%%%\item[]  Collector server:
%%%%%\item 
%%%%%Attack 1: the collector server misleads users by sending them the wrong group information.
%%%%%\item 
%%%%%Attack 2: the collector server puts users with different $D_{0j}$ values into the same group.
%%%%%
%%%%%\vspace{2mm}
%%%%%%\item[]  
%%%%%%\vspace{-1mm}
%%%%%
%%%%%\item[]  User:
%%%%%\item 
%%%%%Attack 3: the user intentionally sends the encrypted index message to the wrong group.
%%%%%\item 
%%%%%Attack 4: the user tampers its own $D_{0j}$ value in the index message.
%%%%%\item 
%%%%%Attack 5: the processor illegally inserts or deletes an encrypted index message.
%%%%%\item 
%%%%%Attack 6: the processor tampers the (encrypted) $D_{0j}$ in an encrypted index message.
%%%%%
%%%%%\vspace{2mm}
%%%%%\item[]  Provider server:
%%%%%\item 
%%%%%Attack 7: the provider server tampers user's data or encrypted index number when submitting to the collector server.

%%%%info2  \item[]  
User (processor):

%%%%info2  \item 
\noindent $\bullet$ Attack 1: the malicious user encrypts the index message using the wrong public key(s), or signs the transmitted messages using wrong signing key.

%%%%info2  \item
\noindent $\bullet$ Attack 2: the malicious user doesn't send its EIM, or sends multiple EIMs to the first IAP processor.

%%%%info2  \item 
\noindent $\bullet$ Attack 3: the malicious IAP processor illegally inserts or deletes EIM(s) during its processing.

%%%%info2  \item 
\noindent $\bullet$ Attack 4: the malicious user doesn't send its DSM, or sends multiple DSMs to the same provider server.

%%%%info2  \item[]  
Provider server:

%%%%info2  \item 
\noindent $\bullet$ Attack 5: the provider server encrypts the DSM to be submitted with the wrong public key or its signature is generated with the wrong signing key. It may also happen that the provider server tampers the signed acknowledgement message of the collector.
%%%%info2  \item 

\noindent $\bullet$ Attack 6: the provider server doesn't submit the DSM, or submits multiple copies of DSMs for a given user.

%%%%info2  \end{itemize}

\begin{figure}[t]
\centering
\begin{minipage}{1.5in}
  \centering
  \includegraphics[width=1.55in, height=2.7in]{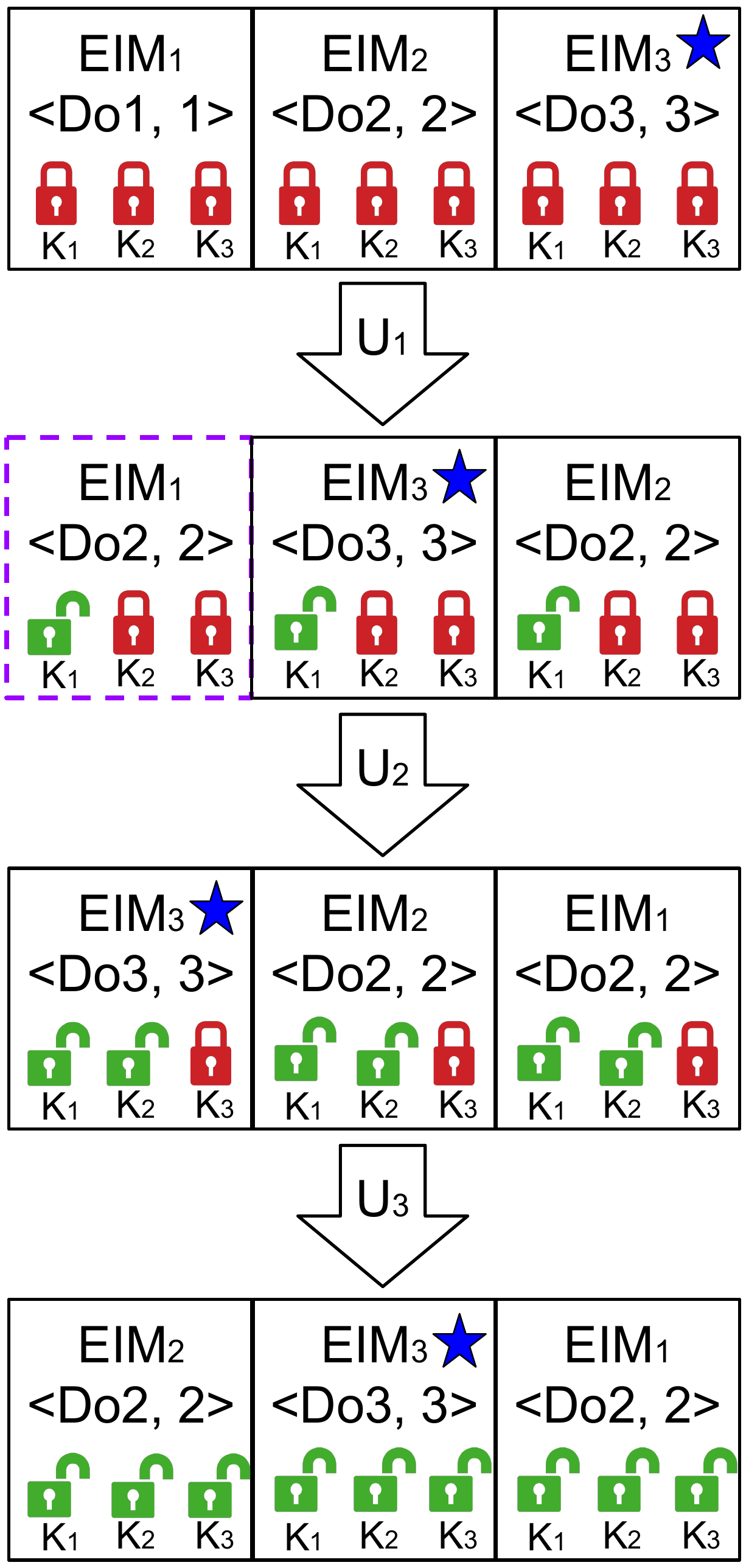}
\vspace{-4mm}
  \caption{The Replication Attack}%{A figure}
  \label{fig:replicate}
\end{minipage}%
\hspace{4mm}
\begin{minipage}{1.5in}
  \centering
  \includegraphics[width=1.55in, height=2.7in]{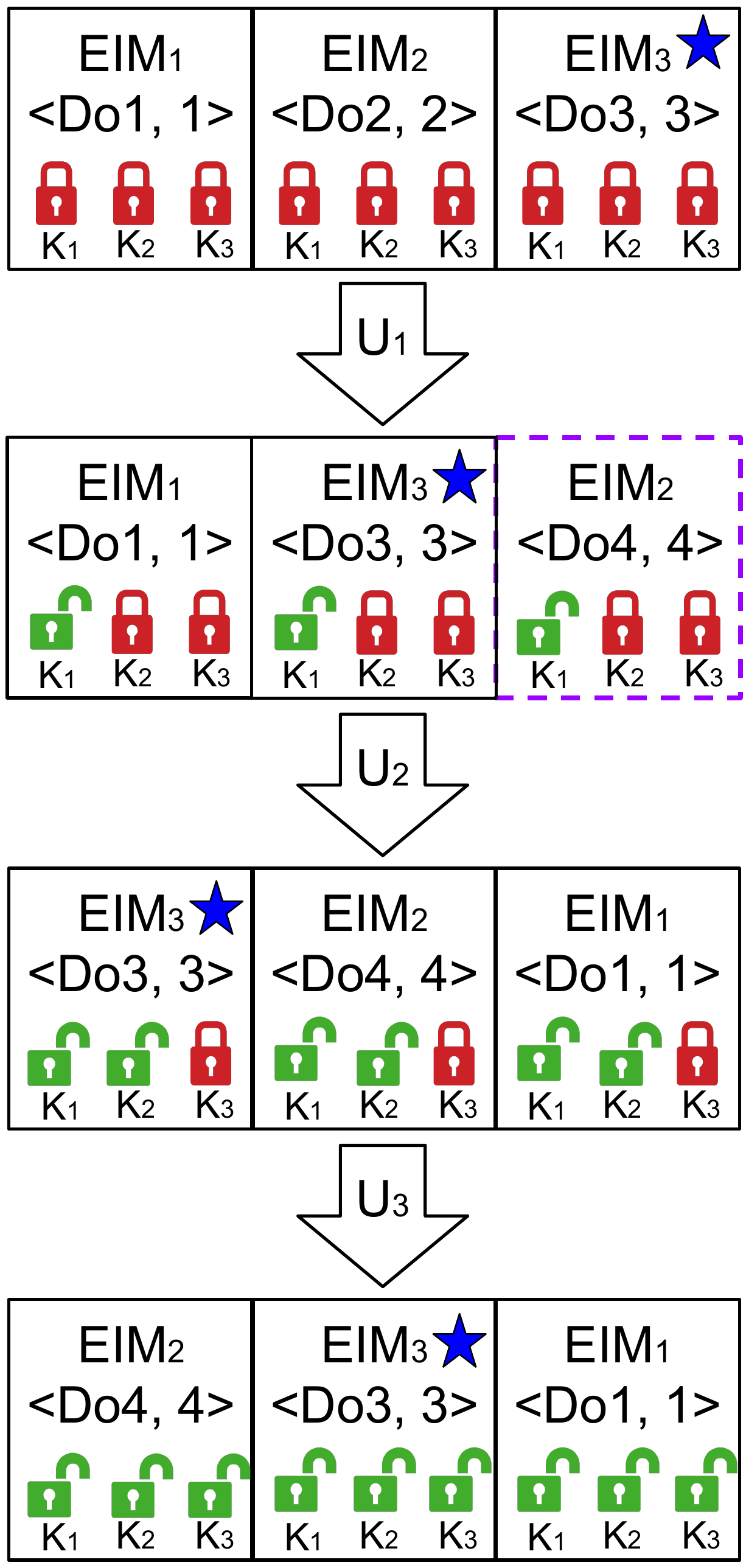}
\vspace{-4mm}
  \caption{The Replacement Attack}%{Another figure}
  \label{fig:replace}
\end{minipage}
\vspace{-4mm}
\end{figure}

\subsubsection{Data Inference Attacks}
%%%%The data inference attack can be launched by a single adversary or by multiple colluding entities. 
We present 4 types of active data inference attacks in which the attacker(s) manipulates the peer-shuffle process to expose the owner identity of the index message(s).
%
%Generally speaking, every user performs shuffle and decryption on all encrypted messages one after another in the peer-shuffle procedure. 
%The unmanipulated peer-shuffle guarantees that if at least two EIMs belong to honest users, attackers are not able to decide the owners of these EIMs after processed by an honest processor.
%%%%%%Info Normally, the attackers are not able to decide the owner of the EIMs that belong to honest users, after processed by an honest processor.
%
From the attacker's perspective, to break the strong anonymization provided by peer-shuffle, the target EIM has to be somehow ``marked" so that it can be recognized even after it has been randomized by honest processors. 
%%%%The replication attack and replacement attack use different techniques to mark the target EIM.
%%%%%
%%%%The unique collector data attack utilizes the uniqueness of $D_{0j}$.
%%%%%
%%%%The inconsistent broadcast attack is like a random guessing.
%
%%%%%%Info For honest users, the exposure of their index messages or numbers may lead to the disclosure of their whole data tuples.

%%%%info2  \begin{itemize}[leftmargin=*]%[nolistsep]
%\vspace{-1mm}

%%%%info2  \item 
\noindent $\bullet$ Attack 7 (replication attack): the compromised users mark the target EIM by replicating other honest users' EIMs (or only the encrypted index number part), so that the target EIM is unique among the honest users. As illustrated in Figure~\ref{fig:replicate}, the three original EIMs contain the index number $1, 2, 3$, respectively. The first processor $U_{1}$ is compromised and it wants to reveal honest user $U_{3}$'s index number. It copies the other honest user $U_{2}$'s $EIM_{2}$ and overwrites its own $EIM_{1}$. Now, it is easy to tell the unique $EIM_{3}$ from the duplicated $EIM_{2}$ in the rest of IAP processes. In the end, the colluding party can obtain $U_{3}$'s specific index number when $EIM_{3}$ is fully decrypted.
There are two prerequisites for a replication attack: (1) the first processor must be compromised so that the originator of the EIMs is known; (2) the number of compromised users must be no less than 1 compared to the number of honest users.

\begin{figure}[t]
%\vspace{-3mm}
\centering
\includegraphics[width=3 in,height=1.1in]{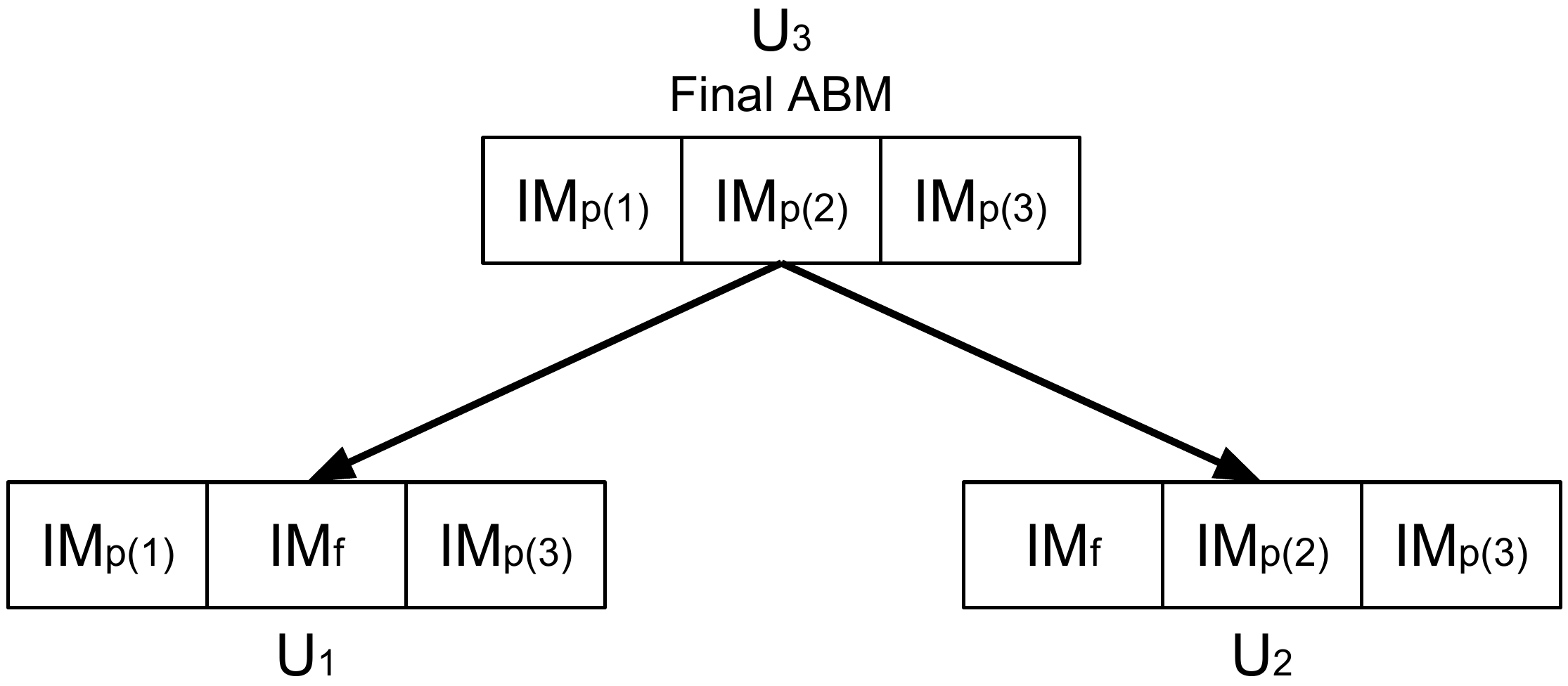}
\vspace{-2mm}
\caption{The Broadcast Attack}
\label{fig:broadcast}
\vspace{-5mm}
\end{figure}

%%%%info2  \item 
\noindent $\bullet$ Attack 8 (replacement attack): the compromised users mark the target EIM by replacing other honest users' EIMs with the fake EIMs they created. The fake EIMs are different to the target EIM and their own EIMs.
In a replacement attack, the colluding entities know exactly what the decrypted fake EIMs look like when they go through each honest processor. 
%%%%info2  Therefore, they are able to track the EIMs of all the rest of honest users except the target user. In other words, the only EIM they do not recognize must belong to the target user.
Therefore, the target EIM is the only one they do not recognize.
As illustrated in Figure~\ref{fig:replace}, the compromised processor $U_{1}$ wants to know honest user $U_{3}$'s index number. During its own IAP process, $U_{1}$ replaces the $EIM_{2}$ of the other honest user $U_{2}$ with a fake EIM. In all the rest IAP processes, $U_{1}$ can identify not only its own $EIM_{1}$, but also the fake $EIM_{2}$ it created. The only unrecognized EIM left would be the target. Eventually, the attackers can obtain the plaintext index number of $U_{3}$.
There is one prerequisite for a replacement attack: the first processor is compromised.
%%%% since only the first processor knows the originator of each EIM.
%
%%%The detection for replacement attack is put at the start of phase 4. 
% It is pointless and even dangerous to perform the detection during the anonymization (phase 3). For instance, some EIMs have been replaced when a certain benign processor executes the IAP. If the manipulated this processor 
%

%%%%info2  \item 
\noindent $\bullet$ Attack 9 (broadcast attack): the last processor is supposed to broadcast the final ABM (i.e., permuted index messages) to all users and the collector server. However, it may infer the index number of honest users by sending them different (manipulated) ABMs. 
As illustrated in Figure~\ref{fig:broadcast}, the last IAP processor $U_{3}$ is malicious while $U_{1}$ and $U_{2}$ are honest. The permutation function $p(j)$ calculates the permuted position of $U_{j}$'s index message. $U_{3}$ is not sure which honest user owns which piece of the two unrecognized index messages. Therefore it creates two manipulated $ABM$s, each has one of the two unknown IMs replaced by the fake index message $IM_{f}$.
This is similar to a guessing attack. If the guess is wrong, both honest users will report that a replacement attack is detected. Otherwise, both of them find their IMs in the manipulated $ABM$s, and $U_{3}$ will know the belonging of the two index messages.
The prerequisite of broadcast attack is that the last processor is malicious.

%%%%info2  \item 
\noindent $\bullet$ Attack 10 (unique collector data attack): the collector server intentionally collects data from $U_{j}$, whose collector data $D_{0j}$ is unique among all users. As a result, its data tuple $T_{j}$ can be identified by the collector server.

%%%%info2  \end{itemize}

\subsubsection{Data Falsification Attacks} 

The data submission messages can be manipulated by the attackers. Here, we consider pure data falsification attacks whose only purpose is to corrupt the aggregated data  (provider data set $P_{j}$).
%%%%The provider data set $P_{j}$ received by the collector may have been falsified by the provider servers.

%%%%info2  \begin{itemize}[leftmargin=*]%[nolistsep]
%\vspace{-0.5mm}

%%%%info2  \item
\noindent $\bullet$ Attack 11: the provider server tampers the DSMs when submitting to the collector server.

%%%%info2  \item
\noindent $\bullet$ Attack 12: 
%a special type of data falsification attack is that 
the malicious user intentionally submits the wrong/falsified data to the collector through honest provider servers. 
%%%%%Note that the collected data could be validated after the data aggregation (e.g. one piece of submitted data is not consistent with others in collaborative sensing). 
This attack can be detected only if the collector has the ability to validate the aggregated data, hence its detection is not included in the protocol as a standard step.

\section{Analysis}

\subsection{Correctness}

%%%%%%Info After a protocol run, every honest user must have either: (1) correctly submitted its data tuple to the collector; (2) realised that the aggregation is aborted. 
When our scheme terminates, either all honest users have correctly submitted their data to the aggregator, or the scheme is aborted and the investigation is launched.
In the first case, all proactive checkings have been passed successfully and the protocol has not been interrupted by any disruption attack. 
The replacement attack checking verifies that the index messages of honest users are in the final ABM. 
The replication attack checking, broadcast consistence checking and uniqueness checking ensure that the index messages have been correctly received by the collector, and their ownerships are not exposed due to replication marking, random guessing or collector data uniqueness.
The data submission checking confirms that the data submission messages has been correctly submitted to the collector.
%%%%%%Info2 , and indexed by the corresponding encrypted index numbers. 
Hence, the collector is able to reconstruct the data tuples by linking the index numbers.
For the second case, the protocol is aborted by a disruption attack or because one of the checkings has failed.

\subsection{Anonymity} 
\label{subsec:passive}

%\vspace{-0.5mm}
Our privacy-preserving aggregation scheme is resistant to a variety of passive data inference attacks. If the proposed protocol is strictly followed, no entity (e.g., user, server, or the global eavesdropper) can passively infer any data that they are not supposed to know, either individually or collaboratively. 
%
%%%%Besides, our scheme can also defend and provide accountability for various active attacks (details presented in Section~\ref{sec:account}).

\subsubsection{Collector server}
users submit their data in two parts. The first part (index messages) is anonymized by peer-shuffle. The second part (data submission messages) is submitted via delegates (i.e., provider servers) as a batch. 
Hence for either of them, the collector server alone cannot trace to the owners.

\subsubsection{Provider server}
each provider server only has access to its own data, and cannot obtain other servers' data by oneself.

\subsubsection{Users}
a user can access each other's data in phase 3. However, peer-shuffle guarantees that the data of honest user will not be exposed unless there is only one honest user.

\subsubsection{Passive collusion between the IoT server and users}
the server (either collector or provider server) may collude with a number of compromised users, to infer the data that does not belong to the colluding entities.
For collusion with the collector server, our scheme is based on peer-shuffle, which can guarantee data anonymity for honest users if there are at least two of them.
% ensure a negligible probability of getting the associations between honest users and their index numbers if there are at least two honest users.
%%%%% provide anonymity between the index numbers and their owners as long as two group members are honest.
%
For collusion with a provider server, the index number is encrypted with random bits in DSM to prevent the leak of collector data $D_{0j}$ to the colluding party. Specifically, colluding users know the connection between $PN_{j}$ and $D_{0j}$ from the broadcasted IMs, while the provider server knows the owner identity of the encrypted index numbers ${E}_{KU_{0}^{S}}^{R_{j}}[ PN_{j} ]$ from the received DSMs. 
The random number $R_{j}$ must be used to break the equality after encryption, so that the colluding party cannot infer the $D_{0j}$ value of honest users by linking the index numbers ($KU_{0}^{S}$ is publicly available).
%%%%%%Info2 Note that the random number $R_{j}$ must be used, otherwise the colluding party will be able to infer the $D_{0j}$ value of honest users by linking the index numbers: if the index numbers are encrypted with $KU_{0}^{S}$ but without padding, the two encrypted numbers can be linked since such encryption does not break the equality characteristic; or if only one of the numbers is encrypted, they can still be matched after the plaintext $PN_{j}$ is encrypted ($KU_{0}^{S}$ is publicly available). 

%\vspace{-0.5mm}
\subsubsection{Passive collusion involving a global eavesdropper}
the passive eavesdropper is capable of monitoring the sender/receiver of all messages. In our scheme, messages are transmitted in phases, with the same length and format within each phase. So the traffic and timing analysis attack by the eavesdropper alone is mitigated. 
However, if it colludes with an entity that owns or can observe the plaintext contained in the encrypted messages, they may work together to de-anonymize the data or the index number in plaintext form.  
In phase 1, the eavesdropper may collude with the compromised users to infer the collector data $D_{0j}$ (e.g. finding if any two are identical). In our scheme, $D_{0j}$ is encrypted with random numbers selected by the collector, which breaks the equality in ciphertexts.
In phase 2 and 3, the peer-shuffle can preserve the privacy of honest users ($\geq 2$) since only each honest user themselves knows the random number used in the serial encryptions.
In phase 4, the eavesdropper may collude with the collector server. Specifically, the eavesdropper knows the originator $U_{j}$ of the encrypted DSMs ${E}_{KU_{i}^{S}}^{R_{j}^{\prime}}[ DSM_{ij} ]$, while the collector server later gets $DSM_{ij}$ from the provider $S_{i}$. Although key $KU_{i}^{S}$ is publicly available, the colluding party cannot associate the $DSM_{ij}$ to the encrypted DSMs in $\Omega_{4.2}$, as they do not know the random number $R_{j}^{\prime}$.
%% reconstruct the encrypted $DSM_{ij}$ in $\Omega_{4.2}$ as they do not know the random number $R_{j}^{\prime}$.
%
Message $\Omega_{5}$ and $\Omega_{6.1}$ are transmitted between servers, so there is no individual user to track.
The $\Omega_{6.2}$ messages contain only the signatures, no data or index number can be exploited.

\subsection{Accountability}
\label{sec:account}

Our scheme can preserve anonymity and provide accountability for the various active attacks launched by the collector, provider, compromised users or multiple colluding entities. 
%%%%%%Info2 Our scheme is able to detect these attacks, and find the responsible attacker (or at least one of the attackers if there are multiple ones who cover for each other). 
If there are multiple attackers who cover for each other, at least one of them can be found.
In general, the investigation is conducted in phases. For each phase, we only need to check if the output is correctly computed from the input message. While if there is an inconsistency with the message transmitted between two entities, either the sender or the receiver is lying. The signed message can prove if the sender is the liar; otherwise, it is the receiver. 
The investigation may need to replay the IAP process. Note that the private keys are not required, the decryption process can be validated by replaying the serial encryptions in reverse using the public keys and the logged random numbers.
Next, we present the detection and investigation for the 12 active attacks listed in Section~\ref{subsec:attack}.

The disruption attacks expose themselves as they break the current protocol run. 
%
%%%Specifically, the disruption attacks break the protocol with abnormalities like missing/extra EIM, decryption failure, invalid signature, extra DSM, etc. For the data inference attacks and the data falsification attacks, we enforce a series of checkings in different phases of the scheme. 
%
Attacks 1 and 5 disrupt the protocol as the receiver of a message cannot successfully decrypt the message or when the signature cannot be verified. The failed decryption can be caused either by faulty encryption or someone has manipulated the message. Similarly, the wrong signature can be caused by either the incorrect signing process or from being tampered with. 
In such cases, the receiver only needs to prove that the problematic messages are indeed coming from the sender (i.e., signed with the sender's signature); otherwise, the receiver is the one to blame. 
If the number of EIMs in phase 3 is not consistent with the number of participating users, the anonymization procedure will be replayed to check if it is due to illegal insertion/deletion by a malicious processor (Attack 3), or a malicious user who does not send or send multiple EIMs to the first processor (Attack 2). The first processor can trace the sender of each EIM, so it knows if anyone is supposed to submit the EIM while did not actually submit. The multiple submission of EIMs can be proved by the multiple signed $\Omega_{2}$ messages from the malicious user.
The missing/extra DSM can be resulted from either a malicious provider server (Attack 6) or a malicious user (Attack 4). The signed $\Omega_{4}$ messages logged between the user and the provider server can prove which one is responsible for the data submission error.

%
%%%%%%Info2 The multiple submission of data by the provider server can be resulted from the malicious provider server (Attack 6) or a malicious user (Attack 4). 
%
%%%%%%Info2 The provider server can prove that a malicious user has submitted multiple copies of data submission requests using the signed $\Omega_{4}$ messages; Otherwise, the provider server itself is the attacker.  

%
The data inference attacks and the data falsification attacks are detected with a series of proactive checkings.
%
%
%In terms of the received messages, 
%When suspicion arises,
%In this manner 
%
The broadcast consistence checking, the uniqueness checking and the data submission checking, if failed, can immediately expose the attacker, which is the last processor (attack 9), the collector (attack 10), and a specific provider server (attack 11), respectively. 
%%%%%%Info2 For the replication attack, the replacement attack and other disruption attacks, the attackers can be found through investigations by replaying the executed procedures and comparing with the logged messages. 
%
The manipulator(s) of EIMs in the replication attack (Attack 7) and the replacement attack (Attack 8) can be found by replaying the IAP processes.
For Attack 12, the investigation of the originator of the falsified data needs to check the specific signed $\Omega_{4}$ message in which it is contained.
%%%%%%Info2 requires the cooperation of the provider servers. Specifically, the provider server who has submitted that piece of data (DSM) needs to show the specific $\Omega_{4}$ message. The signature of that message can expose the malicious sender.
%
%
%
Note that our scheme can detect the data inference attacks and abort the protocol before any provider data is submitted. 
% may expose the index number of honest users through manipulation and collusion
%
%
While for the data falsification attacks, the falsified data can be exposed before finally accepted by the collector.

\subsection{Complexity} 

In this section, we analyze the communication and computational complexity of the proposed scheme. We use $N$ to denote the number of users and $T$ as the number of provider servers participated in a given data aggregation session.
The number of communication rounds, the communication overheads, and the computation overheads will be compared with the prominent peer-shuffle protocol proposed in Dissent \cite{Dissent}. 
%
%%Note that when aggregating data from multiple providers, Dissent has to be modified by adding 
%
Dissent has implemented and evaluated only the ``normal-cases'' of the protocol, we also focus on the efficiency of our scheme assuming all checkings will be successfully passed.
%%%%%%Info 
%%%%as the accountability feature is believed to deter and eventually exclude active attacks. The tracing of the attackers can be viewed as a complementary, while the general performance is decided by the aggregation protocol. 
%%%%In our work, we also focus on the efficiency of the data aggregation and the checkings (we assume all checkings will be passed).

\subsubsection{Number of Communication Rounds} 

The server-user communications in phases 1, 4 ($\Omega_{4.2}$ messages) and 6 ($\Omega_{6.2}$ messages) are parallelizable and require 1 round.
The communications between the collector server and the provider servers in phase 5 and phase 6 ($\Omega_{6.1}$ messages) can be viewed as 1 round.
The submission of EIMs in phase 2 is parallelizable and requires 1 round. Phase 3 cannot be parallelized and requires $N$ rounds. The exchange of hash values in phase 4 ($\Omega_{4.1}$ messages) is parallelizable and requires 1 round.
The total communication rounds for users is $N+5$ (out of the overall communication rounds $N+7$ for the whole IoT system). Dissent \cite{Dissent} considers the traditional one-server-multiple-users architecture, and has $N+4$ communication rounds for users.
%
%%%%Note that this is for the case of single data aggregation (one piece of data per user, i.e. one provider).
%%%%%
%%%%When the collector is aggregating data from multiple providers, only a subset of phases need to be repeated in our scheme. Specifically, the message $\Omega_{4.2}$, $\Omega_{5}$, $\Omega_{6.1}$ and $\Omega_{6.2}$ are sent/received by the additional providers.
%%%%%
%%%%Suppose $T$ providers are involved, the per-user communication rounds of our scheme is $N+2T+3$ (overall there are ??? same), while in Dissent it has increased to $NT+4T$ (??? same)).

\subsubsection{Communication Overheads}

We use $X_{0}$ to denote the unified length for all plaintexts including data, index number, keys, etc. The serial regular public key encryption can keep the ciphertext in fixed-length, while the encryptions with padding will lead to a linear increase in ciphertext length.
We use $X$ to denote the ciphertext length of regular encryptions, and 
$X^{\prime}_{k}$ for the ciphertext length after $k$ rounds of encryptions with padding.
%$X^{\prime}=|C^{\prime}|$ to denote the ciphertext length of the public key encryption with plaintext expansion $C^{\prime}={E}_{KU}^{R}[M]$. 
%
%
The length of hash value and the signature are $H$ and $Y$, respectively. The total length of session ID and phase ID is denoted as $Z$.
%
%Note that the $ABM$s in phase 3 of our scheme and Dissent actually have different $EIM$ length as each processor performs decryption. In our analysis, they are all denoted by symbol $X^{\prime}$ since we mainly focus on the order of $N$ and $T$ in communication complexity.  
%
%
%%%%%%Info The communication overheads of the 6 phases are listed in Table \ref{table:comm}. 
%
%
When aggregating data from $T$ providers, only a subset of phases need to be repeated in our scheme. Specifically, only message $\Omega_{4.2}$, $\Omega_{5}$, $\Omega_{6.1}$ and $\Omega_{6.2}$ are sent/received by the additional $(T-1)$ providers.
%
%%The total communication overhead of single data aggregation (one piece of data per user, i.e. one provider) is $O(N^{2}(2X^{\prime}+Y+Z+H+|D_{0j}|+|PN_{j}|))$, which is about the same level as Dissent's $O(N^{2}(X^{\prime}+X+K+2Y+2Z+H))$ ($K$ is the length of the secondary public keys).
%%%
%%However, when the collector is aggregating data from multiple providers, only a subset of phases involving the providers need to be repeated. The communication overhead increase are the message $\Omega_{4.2}$, $\Omega_{5}$, $\Omega_{6.1}$ and $\Omega_{6.2}$ sent/received by the additional provider, so the overall communication complexity is still $O(N^{2}(2X^{\prime}+Y+Z+H+|D_{0j}|+|PN_{j}|))$. In contrast, the communication overhead of Dissent \cite{Dissent} to collect data from $T$ providers becomes $O(T\cdot N^{2}(X^{\prime}+X+K+2Y+2Z+H))$.
%
The per-user communication overhead of our scheme is $N^{2}(2X_{0}+H+Y+Z)+N\cdot  (\sum_{k=1}^{N} X^{\prime}_{k} + (2T+1)X^{\prime}_{1}+ TX^{\prime}_{2}+X^{\prime}_{N}+ (2T+3)Z + (2T+3)Y)$. In contrast, the per-user communication overhead of Dissent \cite{Dissent} is $T\cdot N^{2}(2X_{0}+H+2Y+2Z)+T \cdot  N \cdot  (\sum_{k=1}^{N} X^{\prime}_{k} + 2Y+2Z-H) -2Y-2Z$, which is approximately $T$ times as much as our scheme.
%
%%%Note that the length of $X^{\prime}$ and $X$ are the same if the expanded plaintext does not exceed the ``maximum plaintext length'; otherwise, the plaintext is splited into two parts for encryption and 
%%%$|X^{\prime}|=2|X|$.
% our scheme N*T: +T\times N(6X^{\prime}+2Y+2Z)
%

\subsubsection{Computation Overheads}

%%%%%%Info The computational overhead of each phase is presented in Table \ref{table:comp}, where 
the $EncptR$, $decptR$, $Sign$, $Veri$, $Hash$ and $SH$ are computations of encryption with padding, decryption of message encrypted with padding, signing function, verification of signature (also the session \& phase ID), hashing function, and shuffle of $N$ objects, respectively. 
%
%%%%%%Info $Match$ is the matching of index number to reconstruct the data tuple.
%%%%, and $Camp$ is the comparison of the data (and index number) actually submitted by the provider to the data submission message generated by the user.
%
%
%
When $T$ providers are involved, the per-user computation overhead of our scheme is $(2N+3T)EncptR + (2N+T+1)DecptR + (T+3)Sign + (N+2T+3)Veri + 2Hash + SH + O(N\log N)$. By comparison, the overhead of Dissent \cite{Dissent} is as large as $NT\cdot EncptR + NT\cdot Encpt + NT\cdot DecptR + 5T\cdot Sign + T(N+3)\cdot Veri + 2T\cdot Hash + T\cdot SH + T\cdot O(N\log N)$, where $Encpt$ is the computation of regular public key encryption.
Note that the complexity of these computations may vary with different length of inputs. 
%%%%We will evaluate the total computaion complexity in terms of execution time via experimental tests.
Overall, Dissent has $O(T)$ times of computation overheads than our scheme.
%
%%%%%%Info2  The serial encryptions/decryptions are the most time-consuming computations, especially those with plaintext padding. 
In the single-provider case, Dissent seems to have lighter per-user encryption overhead $O(N\cdot Encpt+N\cdot EncptR+N\cdot DecptR)$ compared to our scheme's $O(2N\cdot EncptR+2N\cdot DecptR)$.
However, the output of $N$ rounds of $Encpt$ is the input of $N\cdot EncptR$ in Dissent, which means the computation of $N\cdot EncptR$ has to wait until $N\cdot Encpt$ is done. In our scheme, the two sets of $N\cdot EncptR$ are for the collector data and index number, which can be computed in parallel.
%
%

%\vspace{-0.5mm}
\section{Performance Evaluations}

%\vspace{-0.5mm}
We evaluate the performance of our anonymous data aggregation scheme through extensive simulations. All experiments were conducted on an Intel(R) Core(TM) i5 CPU \@ 2.40GHz laptop with 4GB memory. The RSA/ECB/NoPadding and  RSA/ECB/OAEPWithSHA-1AndMGF1Padding cipher transformations in Java cryptography library are used for regular public key encryption and encryption with plaintext-expansion, respectively. 
In our testing program, each entity ran separately, and we counted the total execution time for a successful data aggregation. We also tested with different user group size, data size, and provider group size. The results presented are the average of 20 runs, and are compared with Dissent \cite{Dissent}.
The public/private keys, signing/verification keys, and the secondary public/private keys (used in Dissent) are all 1024 bits. The links between two entities are 5 Mbps with a 100 ms node-to-node latency. Note that our implementation in Java slows down the execution of peer-shuffle compared to \cite{Dissent}, but our main purpose is to demonstrate the relative advantage of our scheme over Dissent.

\begin{figure}[t]
\centerline{
\subfigure[Execution time vs. user group size]{\includegraphics[width=1.7in,height=1.5in]{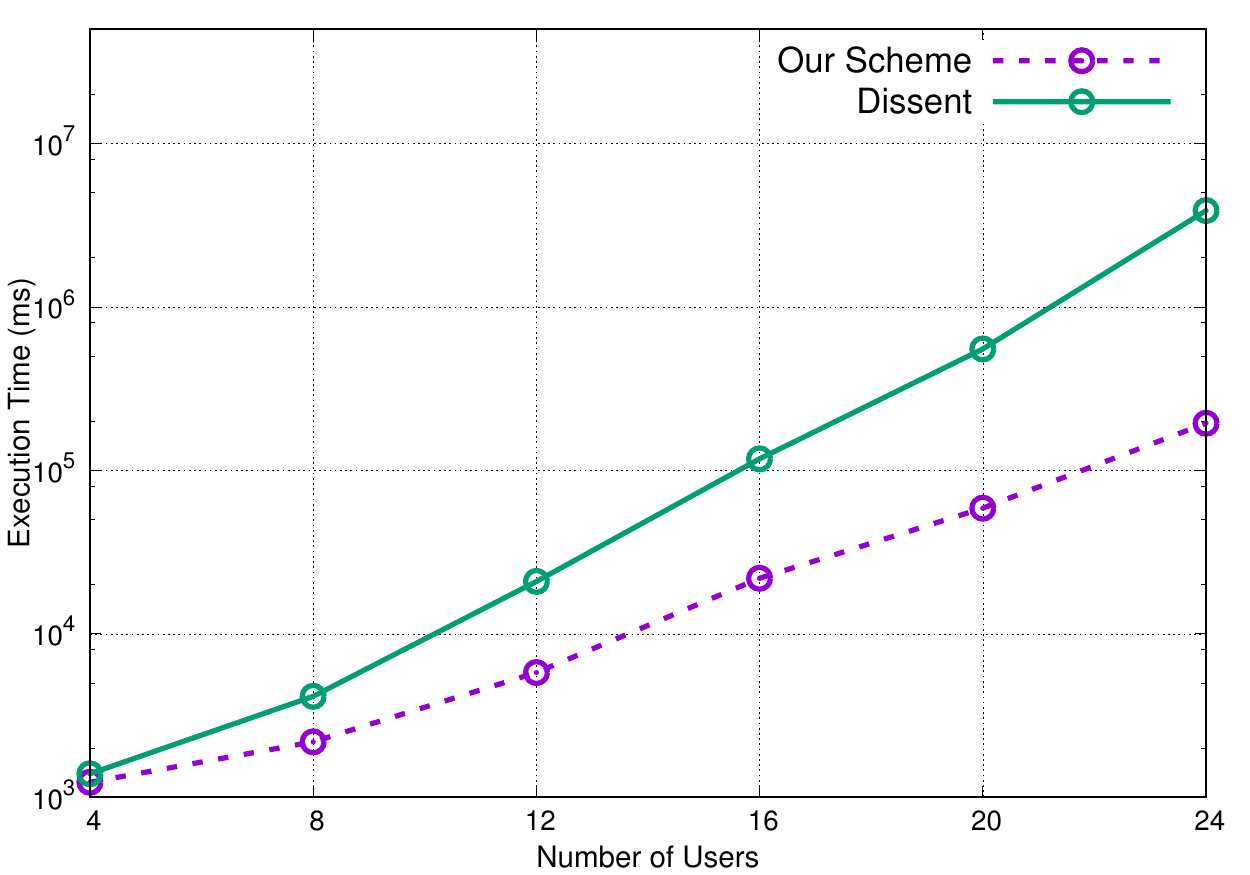}%
\label{fig:VarNumU}} %
%\hfil
\hspace{-1mm}
\subfigure[Communication overhead vs. user group size]{\includegraphics[width=1.7in,height=1.5in]{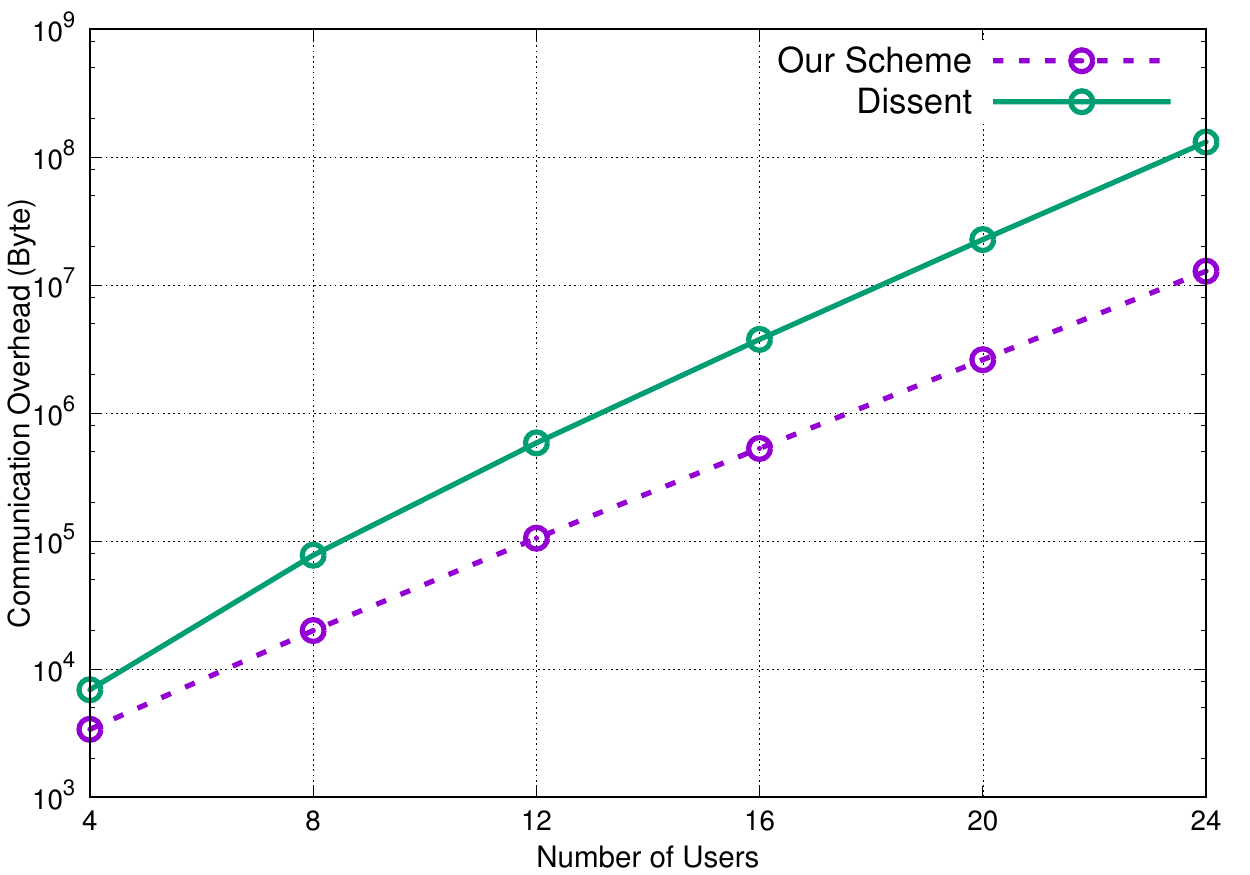}%
\label{fig:CommNumU}} %
%\hfil
\hspace{-1mm}
}
\vspace{-1mm}
\centerline{
\subfigure[Execution time vs. data size]{\includegraphics[width=1.7in,height=1.5in]{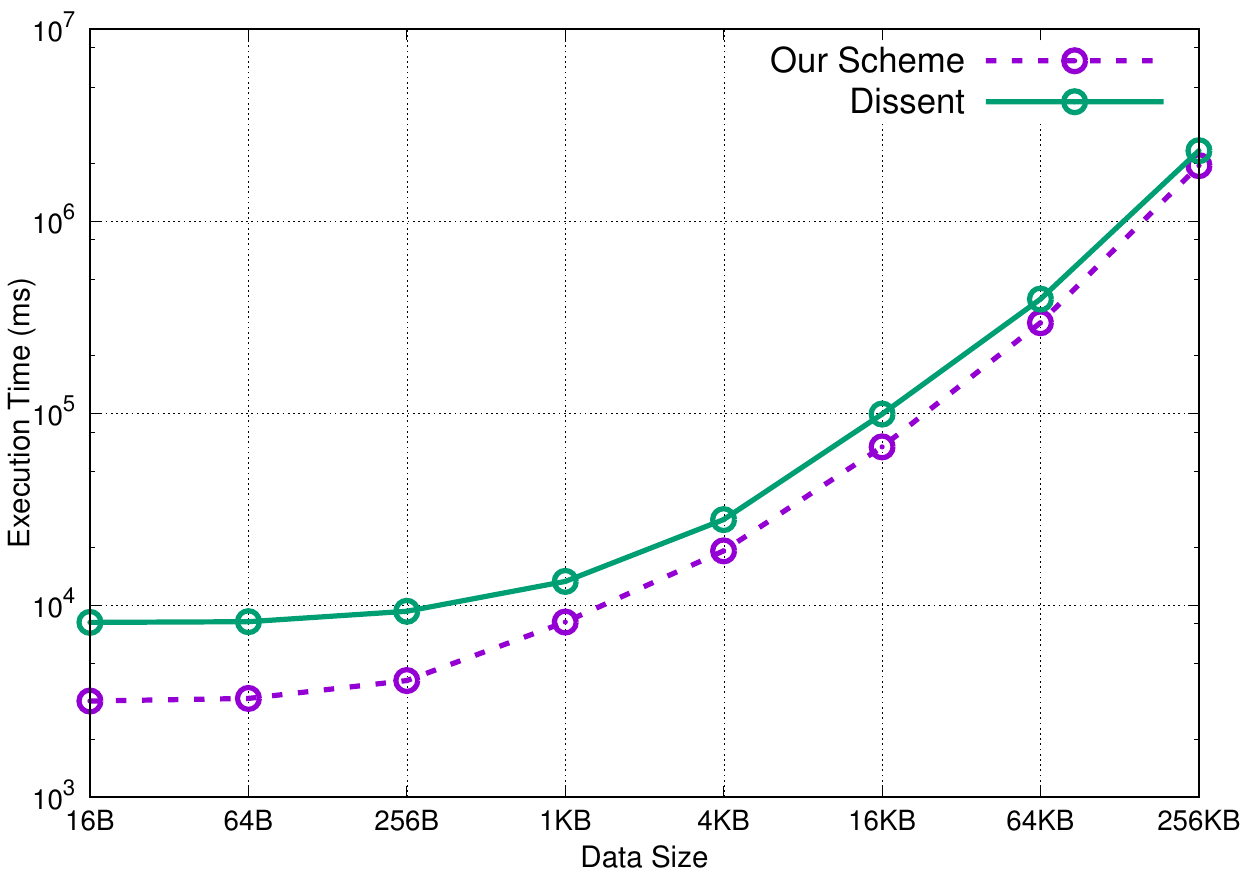}%
\label{fig:VarDataL}} %
%\hfil
\hspace{-1mm}
\subfigure[Communication overhead vs. data size]{\includegraphics[width=1.7in,height=1.5in]{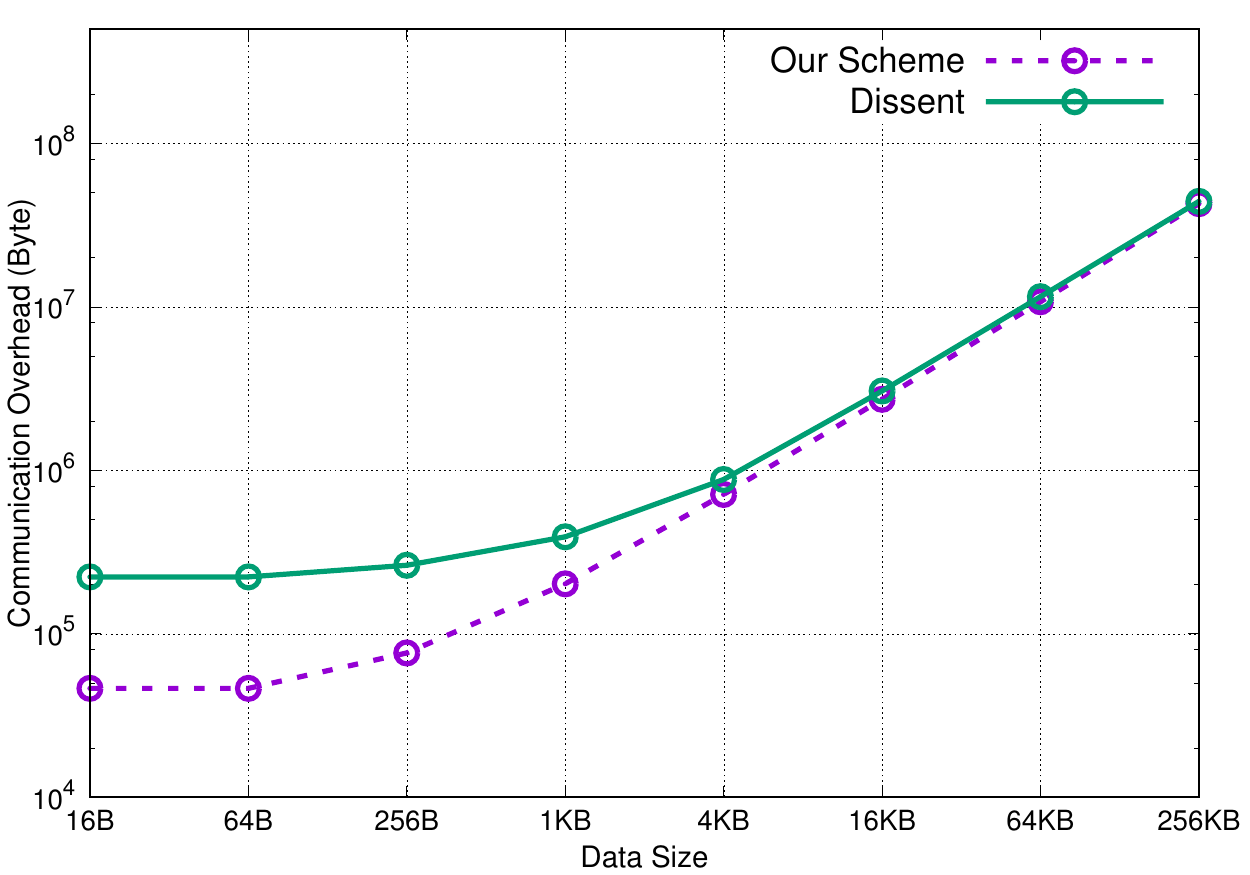}%
\label{fig:CommDataL}} %
}
\vspace{-1mm}
\centerline{
\subfigure[Execution time vs. provider number]{\includegraphics[width=1.7in,height=1.5in]{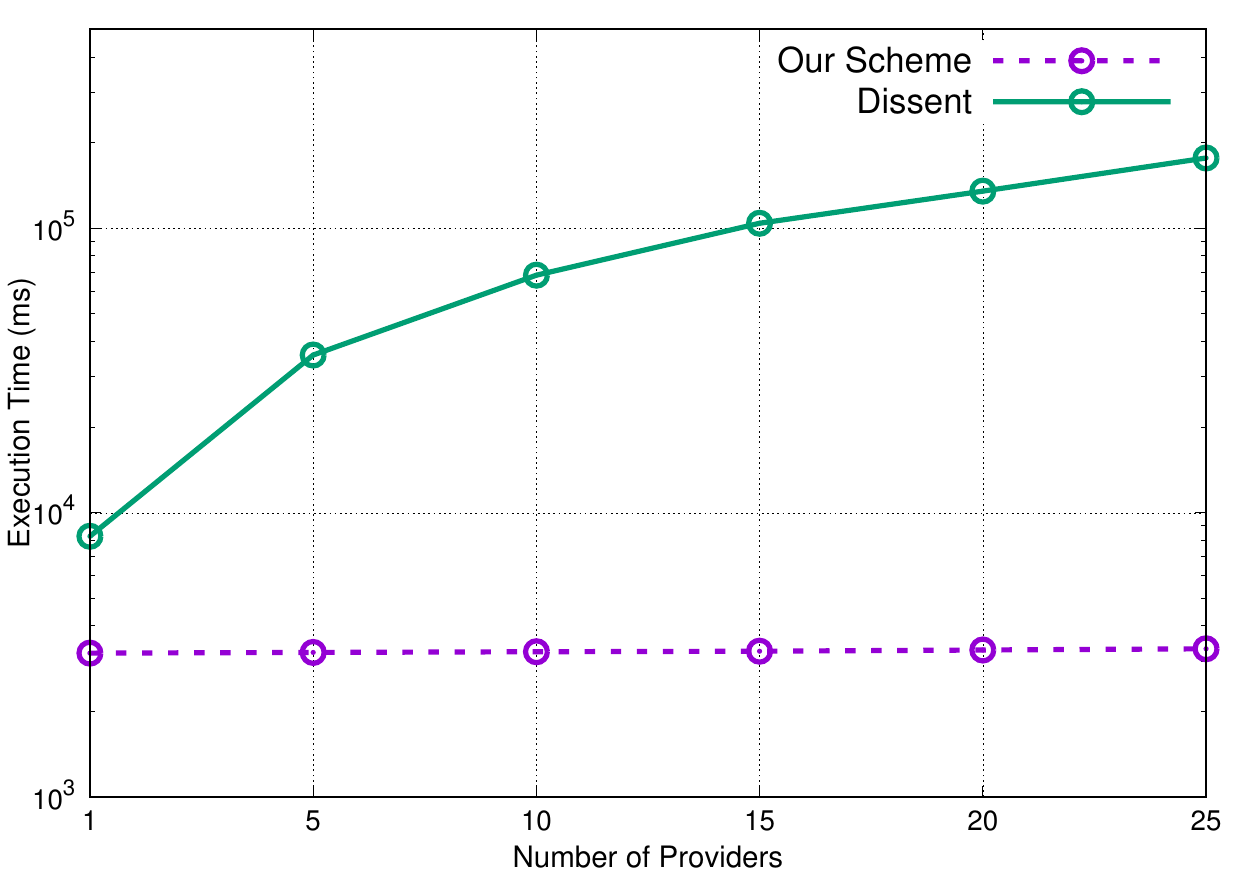}%
\label{fig:VarNumP}} %
%\hfil
\hspace{-1mm}
\subfigure[Communication overhead vs. provider number]{\includegraphics[width=1.7in,height=1.5in]{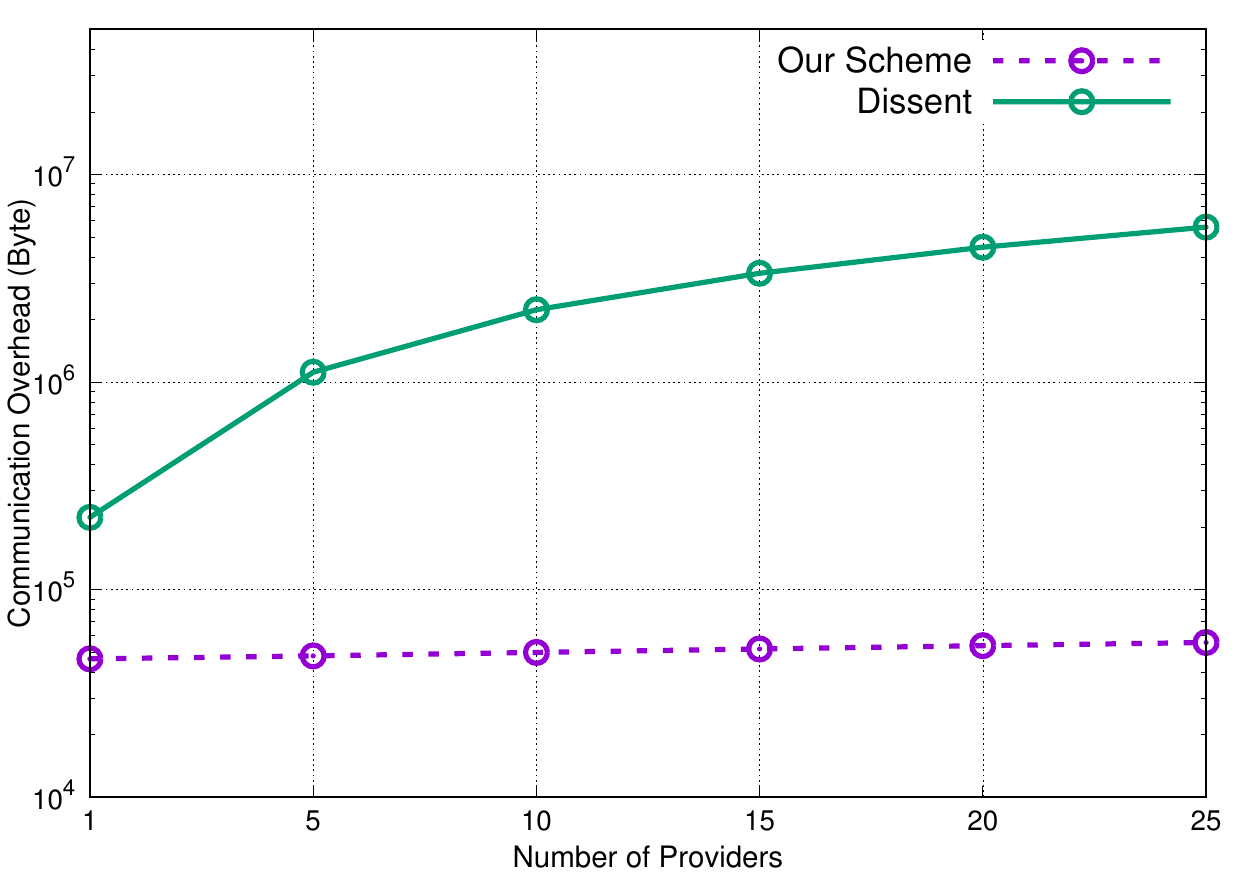}%
\label{fig:CommNumP}} %
}
\vspace{-1mm}
\caption{Performance Evaluation}
\label{fig:evaluation}
\vspace{-4.5mm}
\end{figure}

During implementation we found a practical issue: the plaintext is required to be smaller than the modulus for encryption algorithms like RSA. This can greatly impact the serial encryptions, since the ciphertext generated by a previous round of encryption may exceed the modulus in the next round. 
%
%without padding in Dissent
%Hence in real implementation, we need to limit the input length of regular encryptions to be a little (e.g. 1 byte) smaller than the key length, which 
When it happens, the input has to be split for separate encryptions, which raises the computation workload and the length of the output ciphertext (i.e., extra cipher blocks). 
Dissent and our scheme both perform a set of serial encryptions with padding.
Additionally, the data first go through another set of serial encryptions without padding in Dissent. If such issue occurs in an early stage, the resulted extra cipher blocks will get accumulated in each of the rest serial encryption rounds, triggering increasingly more overheads.
%And this process will be repeated in each round of the serial encryptions (except the first round). The increase in communication overheads can be neglected,  but its impact on the overall computation overheads is huge (it also influences the following $N$ rounds of $EncptR$), as is reflected in our simulation results. Therefore, in practice, our scheme is faster than Dissent even for data aggregation from the single provider.
Although it can be observed from the theoretical analysis that the two schemes have close overheads in single-provider case, the total execution time in Figures~\ref{fig:VarNumU}, \ref{fig:VarDataL} and the per-user communication overhead in Figures~\ref{fig:CommNumU}, \ref{fig:CommDataL} demonstrate that Dissent actually has greater overheads than our scheme.
%
%%%%Info2  When aggregating data from a single provider (provider number is 1), our scheme requires less execution time since it avoids the serial encryptions/decryptions with secondary key pairs used in Dissent.

Figures~\ref{fig:VarNumU} and \ref{fig:CommNumU} present the total execution time and per-user communication overhead when different number of users are involved. The data length is 64 Byte. 
The execution time of both schemes increase with the user group size. Our scheme is more lightweight, especially when the group size is large. Since the serial encryptions/decryptions are the most time-consuming computations for both schemes and a larger user group size means more rounds of serial encryptions/decryptions, the efficiency advantage of our scheme is enlarged as the size of the user group increases.
Similarly, the length of ciphertext data after serial encryptions will also increase with the user group size. Although different processors (users) have different communication overheads, the average value of per-user communication overheads grows larger with every round of encryption that is performed.

Figures~\ref{fig:VarDataL} and \ref{fig:CommDataL} present the total execution time and per-user communication overhead with different data sizes. The user group size is 10.
As we can see, the efficiency merit of our scheme in terms of computation and communication overheads shrinks as the data becomes longer. This is because the serial encryptions with padding hold the dominating part of the overheads when the data size is large, which covers the shortcomings of Dissent including the extra serial encryptions without padding and the aforementioned issue that leads to separate encryptions on plaintext.

%%%%info2  This is because both schemes need to carry out serial encryptions with padding (OAEP \cite{OAEP} is used in our work), which take the dominant part of the computation loads when the data is large.

Besides, our scheme supports the data aggregation from multiple providers, in which the peer-shuffle process including the serial encryptions/decryptions is conducted only once for the index messages. In contrast, the peer-shuffle processing is performed on both the collector data and each set of the provider data in Dissent, which produces huge computation and communication overheads.
Figures~\ref{fig:VarNumP} and \ref{fig:CommNumP} present the performance of our scheme and Dissent when multiple providers are involved. 
The user group size is 10 and the data length is 64 Byte. 
While the total execution time and the per-user communication overheads of Dissent increase with the number of providers, it has little impact on our scheme.

\section{Conclusion}

This paper studies the anonymous data aggregation across multiple companies in the IoT system. This problem cannot be solved properly by previous methods due to its special server-users-servers architecture. We proposed an efficient and accountable anonymous aggregation scheme, which utilizes the semi-trusted provider servers to improve efficiency, and provides resistance and accountability for various attacks. We analyzed and evaluated the communication and computation overheads of our scheme. The experimental results show that our scheme has great efficiency in data aggregation.

\bibliographystyle{abbrv}
\bibliography{./IEEEabrv,./IEEEexample}

\begin{thebibliography}{10}

\bibitem{k-anonymity}
L.~Ahn, A.~Bortz, and N.~J. Hopper.
\newblock K-anonymous message transmission.
\newblock In {\em Proc. of ACM CCS}, 2003.

\bibitem{KDD2006}
J.~Brickell and V.~Shmatikov.
\newblock Efficient anonymity-preserving data collection.
\newblock In {\em Proc. of ACM KDD}, 2006.

\bibitem{mix1981}
D.~Chaum.
\newblock Untraceable electronic mail, return addresses, and digital
  pseudonyms.
\newblock {\em Comm. of the ACM}, 24(2), Feb. 1981.

\bibitem{dc-net1988}
D.~Chaum.
\newblock The dining cryptographers problem: Unconditional sender and recipient
  untraceability.
\newblock {\em Journal of Cryptology}, Mar. 1988.

\bibitem{SS-1}
H.~Corrigan, D.~Boneh, and D.~Mazières.
\newblock Riposte: An anonymous messaging system handling millions of users.
\newblock In {\em Proc. of IEEE S\&P}, 2015.

\bibitem{Dissent}
H.~Corrigan and B.~Ford.
\newblock Dissent: Accountable anonymous group messaging.
\newblock In {\em Proc. of ACM CCS}, 2010.

\bibitem{Verdict}
H.~Corrigan, D.~I. Wolinsky, and B.~Ford.
\newblock Proactively accountable anonymous messaging in verdict.
\newblock In {\em Proc. of USENIX Security}, 2013.

\bibitem{Tor}
R.~Dingledine, N.~Mathewson, and P.~Syverson.
\newblock Tor: The second-generation onion router.
\newblock In {\em Proc. of USENIX Security}, 2004.

\bibitem{management3}
X.~Du, M.~Guizani, Y.~Xiao, and H.~H. Chen.
\newblock A routing-driven elliptic curve cryptography based key management
  scheme for heterogeneous sensor networks.
\newblock {\em IEEE Transactions on Wireless Communications}, 8(3):1223--1229,
  March 2009.

\bibitem{other2}
X.~Du and H.~h.~Chen.
\newblock Security in wireless sensor networks.
\newblock {\em IEEE Wireless Communications}, 15(4):60--66, Aug 2008.

\bibitem{management2}
X.~Du, Y.~Xiao, M.~Guizani, and H.-H. Chen.
\newblock An effective key management scheme for heterogeneous sensor networks.
\newblock {\em Ad Hoc Networks}, 5(1):24--34, 2007.

\bibitem{other5}
X.~Du, M.~Zhang, K.~Nygard, S.~Guizani, and H.-H. Chen.
\newblock Self-healing sensor networks with distributed decision making.
\newblock {\em International Journal of Sensor Networks}, 2(5-6):289--298,
  2007.

\bibitem{Elgamal}
T.~El~Gamal.
\newblock A public key cryptosystem and a signature scheme based on discrete
  logarithms.
\newblock In {\em Proc. of CRYPTO}, pages 10--18, 1985.

\bibitem{dc-net-3}
S.~Goel, M.~Robson, M.~Polte, and E.~G. Sirer.
\newblock {Herbivore: A Scalable and Efficient Protocol for Anonymous
  Communication}.
\newblock Technical report, Cornell University, February 2003.

\bibitem{GuanIoTJ}
Z.~Guan, J.~Li, L.~Wu, Y.~Zhang, J.~Wu, and X.~Du.
\newblock Achieving efficient and secure data acquisition for cloud-supported
  internet of things in smart grid.
\newblock {\em IEEE Internet of Things Journal}, 4(6):1934--1944, Dec 2017.

\bibitem{GuanCommMaga}
Z.~Guan, G.~Si, X.~Zhang, L.~Wu, N.~Guizani, X.~Du, and Y.~Ma.
\newblock Privacy-preserving and efficient aggregation based on blockchain for
  power grid communications in smart communities.
\newblock {\em IEEE Communications Magazine}, 56(7):1--7, Jul 2018.

\bibitem{other1}
X.~Hei and X.~Du.
\newblock Biometric-based two-level secure access control for implantable
  medical devices during emergencies.
\newblock In {\em Proceeding of IEEE INFOCOM}, April 2011.

\bibitem{other3}
S.~Liang and X.~Du.
\newblock Permission-combination-based scheme for android mobile malware
  detection.
\newblock In {\em 2014 IEEE International Conference on Communications (ICC)},
  pages 2301--2306, June 2014.

\bibitem{RAC}
S.~B. Mokhtar, G.~Berthou, A.~Diarra, V.~Quéma, and A.~Shoker.
\newblock Rac: A freerider-resilient, scalable, anonymous communication
  protocol.
\newblock In {\em Proc. of IEEE ICDCS}, July 2013.

\bibitem{OAEP}
{Optimal Asymmetric Encryption Padding}.

\bibitem{OR1998}
M.~G. Reed, P.~F. Syverson, and D.~M. Goldschlag.
\newblock Anonymous connections and onion routing.
\newblock {\em IEEE Journal on Selected Areas in Communications}, May 1998.

\bibitem{Crowds}
M.~K. Reiter and A.~D. Rubin.
\newblock Crowds: Anonymity for web transactions.
\newblock {\em ACM Transactions on Information and System Security (TISSEC)},
  Nov 1998.

\bibitem{OSDI2012}
D.~Wolinsky, H.~Corrigan, B.~Ford, and A.~Johnson.
\newblock Dissent in numbers: Making strong anonymity scale.
\newblock In {\em Proc. of USENIX OSDI}, 2012.

\bibitem{other4}
L.~Wu, X.~Du, and X.~Fu.
\newblock Security threats to mobile multimedia applications: Camera-based
  attacks on mobile phones.
\newblock {\em IEEE Communications Magazine}, 52(3):80--87, March 2014.

\bibitem{management1}
Y.~Xiao, V.~K. Rayi, B.~Sun, X.~Du, F.~Hu, and M.~Galloway.
\newblock A survey of key management schemes in wireless sensor networks.
\newblock {\em Computer Communications}, 30(11):2314 -- 2341, 2007.
\newblock Special issue on security on wireless ad hoc and sensor networks.

\bibitem{IoT}
Y.~Yao, L.~T. Yang, and N.~N. Xiong.
\newblock Anonymity-based privacy-preserving data reporting for participatory
  sensing.
\newblock {\em IEEE Internet of Things Journal}, Oct 2015.

\bibitem{SS-2}
X.~Zhao, L.~Li, G.~Xue, and G.~Silva.
\newblock Efficient anonymous message submission.
\newblock In {\em Proc. of IEEE INFOCOM}, March 2012.

\end{thebibliography}

\end{document}